\documentclass[aps,pra,superscriptaddress,showpacs]{revtex4}

\usepackage{color}
\usepackage{graphicx}
\usepackage{rotating}
\usepackage{amsmath}
\usepackage[a4paper, width=14cm]{geometry}

\begin{document}

\newcommand{\mycomment}[1]{\textcolor{red}{#1}}
\newcommand{\bra}[1]{\langle #1|}
\newcommand{\ket}[1]{|#1\rangle}
\newcommand{\braket}[2]{\langle #1|#2\rangle}
\newcommand{\expectation}[1]{\langle #1 \rangle}
\newcommand{\creatop}[1]{\hat{a}^{\dagger}_{#1}}
\newcommand{\anniop}[1]{\hat{a}_{#1}}

\title[Conditional preparation of pure single photons]{Conditional preparation of single photons using parametric downconversion: A recipe for purity}

\author{P J Mosley, J S Lundeen, B J Smith and I A Walmsley}

\address{Clarendon Laboratory, University of Oxford, Parks Road, Oxford, OX1 3PU, UK}

\email{p.mosley1@physics.ox.ac.uk}


\begin{abstract}
In an experiment reported recently [\textit{Phys.~Rev.~Lett.}, \textbf{100}, 133601, (2008)], we demonstrated that, through group velocity matched parametric downconversion, heralded single photons can be generated in pure quantum states without spectral filtering. The technique relies on factorable photon pair production, initially developed theoretically in the strict collinear regime; focusing --- required in any experimental implementation --- can ruin this factorability. Here we present the numerical model used to design our single photon sources and minimize spectral correlations in the light of such experimental considerations. Furthermore, we show that the results of our model are in good agreement with measurements made on the photon pairs and give a detailed description of the exact requirements for constructing this type of source.
\end{abstract}

\maketitle

\section{Introduction}
\label{sec:intro}

Single photons are one of the most fundamental entities in physics. In recent years, the ability to consistently create and manipulate both single photons and pairs of entangled photons has facilitated everything from tests of quantum theory to the implementation of enhanced precision measurements. High-quality single photons are the resource required for optical implementations of quantum information processing (QIP), yet preparing high purity single photons demands a level of control over all the photonic degrees of freedom that is not simple to achieve.

In a recent paper \cite{Mosley2008Heralded-Generation-of-Ultrafast}, we reported the production of heralded single photons in pure quantum states directly from a parametric downconversion (PDC) source. Due to the relative simplicity of the apparatus required, photon pair generation by PDC in a $\chi^{(2)}$ nonlinear crystal has been the preeminent method of preparing single photons, each conditioned on the detection of its twin. Downconversion is governed by energy and momentum conservation between the pump and daughter fields so that in general the properties of one photon are highly dependent on those of its twin. The degree of intrapair correlation is set by the type of nonlinear crystal and the bandwidth of the pump beam; usually no control is exerted over the modal structure of the photons at the point of their creation and the output is highly multimode with strong correlations in all degrees of freedom.

These correlations within each pair result in mixedness in the heralded single photons. In order to see high-visibility Hong--Ou--Mandel interference (HOMI) between two single photons from independent sources, one must ensure that they are both pure and indistinguishable. If the photons originate in correlated pairs emitted into more than one pair of spatio-temporal modes then the act of heralding, and thereby tracing over the degrees of freedom of the herald photon, sums incoherently over the modes in which the remaining single photon can be found. This mixed state is then of little use.

To date, experiments incorporating PDC-based photon sources have been forced to rely on severe spectral and spatial filtering to restrict the collection of photon pairs solely to those in useable spatio-temporal modes. By filtering towards a single mode, the goal of unit purity becomes obtainable but only in the limit of zero filter bandwidth. This process of filtering discards the majority of the output from the downconverter, drastically reducing the heralding efficiency of such sources. This becomes a more serious problem as one moves towards performing experiments with higher photon numbers.

In this paper, we give details about our PDC source of exceptionally pure photons. By exploiting the technique of group velocity matching between an ultrashort pump pulse and one daughter photon, we can limit the spectral modes into which pairs are emitted and thus produce pure heralded single photons directly from a downconversion source with no spectral filtering, as reported in Reference \cite{Mosley2008Heralded-Generation-of-Ultrafast}. We discuss the factors that affect the purity of the heralded photons and how one can model numerically such a source. Furthermore, we outline the experimental parameters required to yield pure photons.

\subsection{Background and theory}

Here we consider only the effects of frequency correlations on the purity of the resulting heralded single photons; the downconversion is restricted to being approximately collinear and coupled into single mode fibres to select only a single transverse spatial mode. Downconversion takes place in a type-II phasematched nonlinear crystal; the daughter photons are labelled e and o, denoting extraordinary and ordinary polarisations, and are derived from a pump beam labelled p. The frequency dependence of the general state from a downconverter, up to the two-photon component and ignoring the vacuum component which will anyway be eliminated by heralding, can then be written
\begin{equation}
\label{eq:factorability_1}
\ket{\Psi} = \int_{0}^{\infty} d\omega_{e} \int_{0}^{\infty} d\omega_{o} f(\omega_{e}, \omega_{o}) \creatop{e}(\omega_{e})  \creatop{o}(\omega_{o}) \ket{0},
\end{equation}
where $\creatop{e}(\omega_{e})$ is the creation operator for a photon of angular frequency $\omega_e$ in polarisation mode e, and similarly for $\creatop{o}(\omega_{o})$. The joint spectral amplitude $f(\omega_{e}, \omega_{o})$ of the photon pair is defined as the product of the pump envelope function $\alpha(\omega_{e} + \omega_{o})$ and the crystal's phasematching function, $\phi(\omega_{e}, \omega_{o})$:
\begin{equation}
\label{eq:factorability_1b}
f(\omega_{e}, \omega_{o}) = \alpha(\omega_{e} + \omega_{o}) \phi(\omega_{e}, \omega_{o}),
\end{equation}
and the joint spectral probability distribution is given by $\left| f(\omega_{e}, \omega_{o}) \right|^2$. Equation (\ref{eq:factorability_1}) clearly represents a pure state, yet the joint spectral amplitude generally contains correlations between the frequencies of the daughter photons. As a result of this combination of purity and correlation, $\ket{\Psi}$ is entangled in the frequency of the signal and idler photons.

The purity of either heralded single photon derived from $\ket{\Psi}$ can be predicted from the two reduced density operators for the daughter photons. These reduced density operators are
\begin{equation}
\label{eq:matrix_1}
\hat{\rho}_e = \text{Tr}_o \hat{\rho}, \hspace{0.5cm} \hat{\rho}_o = \text{Tr}_e \hat{\rho}
\end{equation}
where $\text{Tr}_i$, is the partial trace over the subsystem $i = e,o$, and $\hat{\rho} = \ket{\Psi} \bra{\Psi}$. The purity of the individual photons is then defined as
\begin{equation}
\label{eq:matrix_2}
\mathcal{P}_e = \text{Tr} (\hat{\rho}_e^2), \hspace{0.5cm} \mathcal{P}_o = \text{Tr} (\hat{\rho}_o^2).
\end{equation}
To see how spectral correlations in $\ket{\Psi}$ affect the purity of the heralded single photons, we consider the Schmidt decomposition of the joint two-photon state \cite{Law2000Continuous-Frequency-Entanglement:, Chan2003Quantum-entanglement-in-photon-atom}. The Schmidt decomposition is a unique method of expressing a bipartite system in terms of a complete set of basis states; under this transformation (\ref{eq:factorability_1}) becomes
\begin{equation}
\label{eq:schmidt_1}
\ket{\Psi} = \sum_{j}\sqrt{\lambda_{j}}\ket{\zeta_{e,j}} \ket{\xi_{o,j}},\hspace{0.5cm} \sum_{j}\lambda_{j} = 1.
\end{equation}
The orthonormal basis states $\ket{\zeta_{e,j}}$ and $\ket{\xi_{o,j}}$ are known as Schmidt modes; each set is dependent on only one subsystem of $\ket{\Psi}$. Each pair of modes is weighted by its Schmidt magnitude, $\lambda_j$. The number of elements required in the sum to express $\ket{\Psi}$ in terms of its Schmidt modes indicates the degree of factorability of the two-photon state. This can be quantified by the Schmidt number, $K$, defined as \cite{Grobe1994Measure-of-electron-electron-correlation, Eberly2006Schmidt-analysis-of-pure-state}
\begin{equation}
\label{eq:schmidt_2}
K \equiv \frac{1} {\sum_{j}{\lambda_{j}}^{2}} \equiv \frac{1} {\textrm{Tr} (\hat{\rho}_{e}^{2})} \equiv \frac{1} {\textrm{Tr} (\hat{\rho}_{o}^{2})}.
\end{equation}
$K$ indicates how many frequency Schmidt modes are active in the two-photon state and hence it is an entanglement measure. A product state will be unentangled and can be described using only one pair of Schmidt modes; therefore it will have $\lambda_{0} = 1, \lambda_{j\geq1} = 0$ and $K$ equal to unity. A state maximally entangled in frequency would require an infinite number of Schmidt modes to describe it, each with a vanishingly small $\lambda_{j}$ coefficient, and therefore $K$ would equal infinity. Comparing (\ref{eq:matrix_2}) and (\ref{eq:schmidt_2}) we see that the purity of both reduced states is equal to the sum of the squares of the Schmidt coefficients:
\begin{equation}
\label{eq:schmidt_3}
\mathcal{P}_e = \mathcal{P}_o = \sum_{j}{\lambda_{j}}^{2}.
\end{equation}
Therefore, to obtain heralded single photons in pure states, one must ensure that only a single Schmidt mode is active so that $\lambda_{0} = 1$ and $\mathcal{P}_e = \mathcal{P}_o = 1$. From (\ref{eq:schmidt_1}) it is clear that when this criterion is fulfilled, $\ket{\Psi}$ is a product state
\begin{equation}
\label{eq:schmidt_4}
\ket{\Psi} = \ket{\zeta_{e}}\ket{\xi_{o}} = \int_{0}^{\infty} d\omega_{e} f_e(\omega_{e}) \creatop{e}(\omega_{e}) \int_{0}^{\infty} d\omega_{o} f_o(\omega_{o})  \creatop{o}(\omega_{o}) \ket{0},
\end{equation}
and the joint spectral amplitude is factorable into a function of $\omega_e$ only multiplied by a function of $\omega_o$ only: $f(\omega_{e}, \omega_{o}) = f_e(\omega_{e}) f_o(\omega_{o})$. Hence the only way in which the heralded photons can remain in a pure state is if there is no frequency correlation in the initial state; this can only be realized by making $f(\omega_{s}, \omega_{i})$ factorable. Detecting one photon then has no effect on the other as no distinguishing information about the properties of the twin is available.

\subsection{The singular value decomposition}

For a specific set of two-photon states (such as those considered in references \cite{Grice2001Eliminating-frequency-and-space-time, URen2005Generation-of-pure-state-single-photon}), it is possible to find analytically the Schmidt decomposition (\ref{eq:schmidt_1}) of a good approximation to the state. However, the joint states considered in this paper do not fall within this set and an analytic method cannot be used to find the anticipated purity of heralded photons from the joint distributions that we will be concerned with. Hence it would be convenient to have a method of finding the anticipated purity of conditionally prepared single photons from any arbitrary joint amplitude \cite{Lamata2005Dealing-with-entanglement}. Fortunately, there exists a matrix operation, known as the singular value decomposition (SVD), that is the analog of the Schmidt decomposition and can be computed numerically for any input state.

Let the state $\ket{\Psi}$ be represented by the square matrix $F$ where $F_{mn}$ is the matrix element representing $f(\omega_{e,m}, \omega_{o,n})$ and $\omega_{e,m}$ and $\omega_{o,n}$ are discrete frequency components of the e-ray and o-ray respectively. The SVD of $F$ is then defined as the decomposition of $F$ into three matrices, two of which are unitary, $U$ and $V^{\dag}$, and one diagonal, $D$, such that
\begin{equation}
\label{eq:svd_1}
F = U D V^{\dag}.
\end{equation}
The unitary matrices then contain the modes into which the initial state has been decomposed. $U$ is dependent only on $\omega_e$ and its $j^{\mathrm{th}}$ column, $U_{mj}$ (where $j$ is held constant over e-ray frequency index $m$), represents the e-ray Schmidt mode $\ket{\zeta_{e,j}}$ while $V^{\dag}$ depends only on $\omega_o$ and the $j^{\mathrm{th}}$ row $V^{\dag}_{jn}$ describes the o-ray Schmidt mode $\bra{\xi_{o,j}}$. The diagonal elements of $D$ are called the singular values of $F$; they are non-negative and appear in descending order of magnitude. For a normalized state $\ket{\Psi}$ these singular values are identical to the Schmidt magnitudes. This can be seen by multiplying out the three matrices on the right hand side of (\ref{eq:svd_1}); each element of $D$ describes how much of each of the modes listed in $U$ and $V^{\dag}$ are mixed at each pair of e-ray and o-ray frequencies to form $F$. Therefore the elements of $F$, specifying the amplitude at every pair of frequencies, each contain a sum over all $j$ mode pairs.

The singular values can be swiftly calculated numerically for relatively large matrices and the Schmidt number found simply by normalizing $D$ and summing the squares of the elements. This technique is particularly powerful as the Schmidt number can be determined quickly for states for which an analytic solution to the Schmidt decomposition does not exist.

\section{Factorable state generation in KDP}

The theoretical framework for the generation of factorable two-photon states in the strictly collinear regime was expounded by Grice \textit{et al} \cite{Grice2001Eliminating-frequency-and-space-time} and U'Ren \textit{et al} \cite{URen2003Managing-photons-for-quantum, URen2005Generation-of-pure-state-single-photon}. The underlying principle is to control the modes into which photon pairs can be emitted and thus generate pairs that have no frequency correlations and are therefore factorable. If the photon pairs can be generated in only one joint spatio-temporal mode, $\ket{\Psi}$ can be decomposed into a single pair of Schmidt modes, fulfilling the factorability criterion. Approximating the collinear phasematching function as a Gaussian distribution and expanding the wavevector mismatch only as far as the second order, it can be shown that the condition that must be fulfilled to attain factorable pairs is \cite{URen2005Generation-of-pure-state-single-photon}:
\begin{equation}
\gamma \frac{L^2}{2} \left(\frac{\partial k_p}{\partial \omega} - \frac{\partial k_e}{\partial \omega} \right) \left( \frac{\partial k_p}{\partial \omega} - \frac{\partial k_o}{\partial \omega} \right) + \frac{2}{\sigma^2} = 0,
\label{eq:condition1}
\end{equation}
where $\gamma = 0.193$ matches the width of the Gaussian approximation to the phasematching function with that of the genuine sinc phasematching function, $L$ is the crystal length, $\sigma$ the pump bandwidth, and $\partial \omega / \partial k_j = v_{g,j}$, the group velocities of the three fields. Rewriting the wavevector derivatives as mismatches between the inverse group velocities of pump and e-ray, $\Delta v_e^{-1}$, and pump and o-ray, $\Delta v_o^{-1}$, we see that
\begin{equation}
\gamma \sigma L \Delta v_o^{-1} + \frac{4}{\sigma L \Delta v_e^{-1}} = 0.
\label{eq:condition2}
\end{equation}
Here we satisfy this condition through two constraints. Firstly, the group velocity of one daughter photon (the o-ray) must be the same as that of the pump so that the first term in (\ref{eq:condition2}) is zero. For PDC this condition can be met in  any nonlinear crystal for particular sets of pump and downconverted wavelengths, though for most common crystals the wavelength of at least one of the daughter photons will be outside the range of silicon photodetectors (greater than 1\,$\mu$m). Secondly, the product of the pump bandwidth and crystal length must be large enough relative to the inverse group velocity mismatch of the e-ray photon that the second term in (\ref{eq:condition2}) is negligibly small, i.e.\,$\sigma L \gg \Delta v_e$. This can be expressed in terms of the transit time difference through the crystal of the pump and e-ray, $\Delta \tau_e = L \Delta v_e^{-1}$, and we have the equivalent condition that the inverse pump bandwidth must be much smaller than the e-ray transit time difference: $\sigma^{-1} \ll \Delta \tau_e$.

It is worth noting here that this scheme is dependent upon having a pump pulse with a duration that is approximately at the Fourier transform limit when it reaches the centre of the PDC crystal. While any temporal chirp on the pump pulse does not introduce additional spectral correlations it will give rise to temporal correlations that reduce the purity of the heralded state in just the same way, though in the long crystal limit this effect is small \cite{URen2005Generation-of-pure-state-single-photon}. In the analysis contained herein it is assumed that the pump pulses are transform limited.

\begin{figure}
\begin{center}
\includegraphics[width= 0.65 \textwidth]{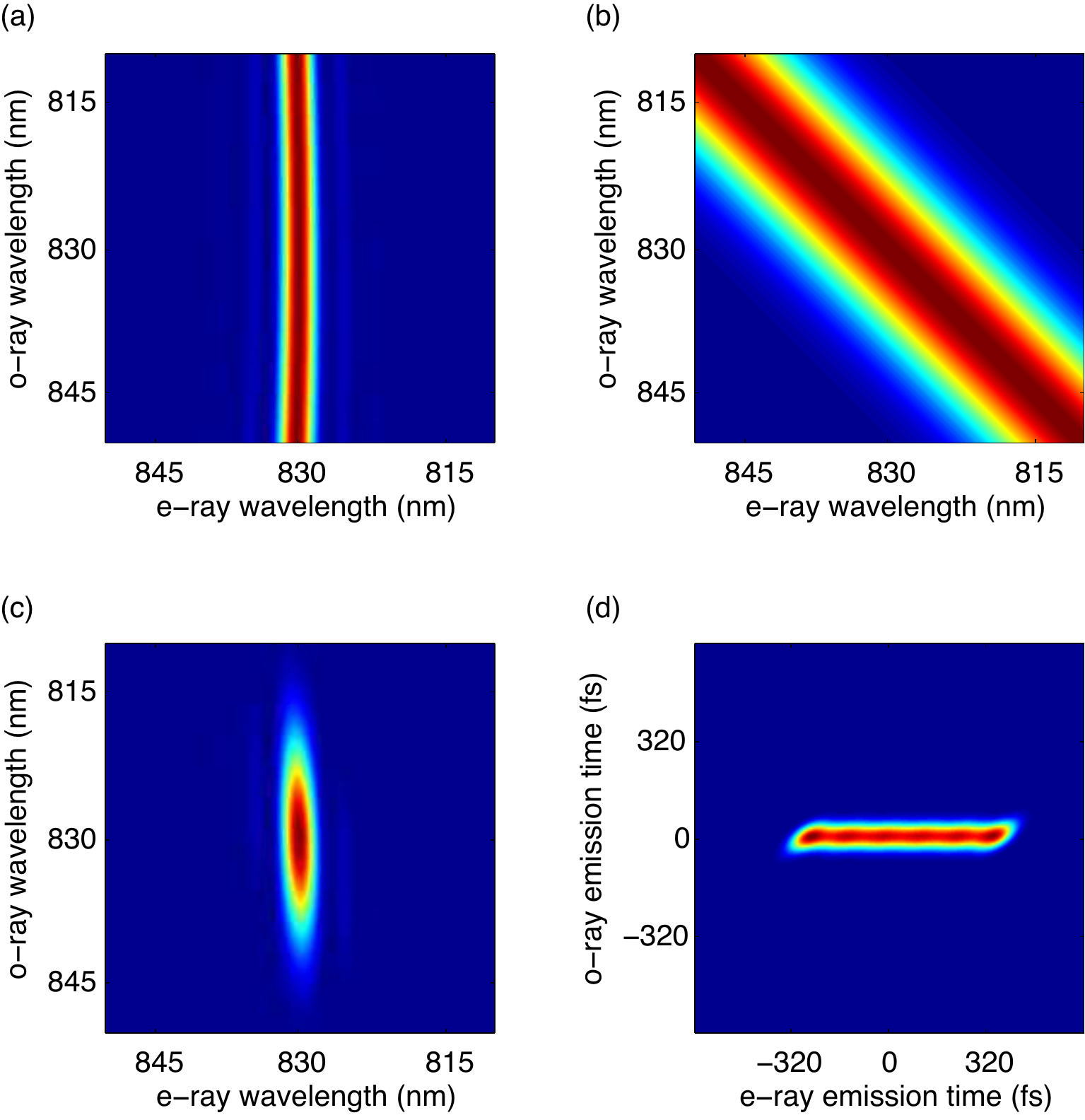}
\caption{Collinear phasematching intensity (a), pump envelope intensity (b), joint spectral intensity (c), and joint temporal intensity (d) for a 5\,mm KDP crystal pumped at 415\,nm with 4\,nm FWHM bandwidth pulses.}
\label{fig:kdp_jsi}
\end{center}
\end{figure}

Here we study the particular case of PDC in a potassium-dihydrogen-phosphate (KDP) crystal. This negative uniaxial crystal was chosen because for a beam propagating at an angle of 67.8$^{\circ}$ to the optic axis (the phasematching angle for production of collinear downconversion at 830\,nm), an e-polarised pulse at 415\,nm will travel with the same group velocity as an o-ray pulse at 830\,nm. Hence the group velocity matching condition is satisfied for pair generation at a wavelength for which silicon avalanche photodiodes (APDs) provide highly efficient single photon counting and the nonlinear crystals can be pumped by a frequency-doubled femtosecond Ti:Sapphire oscillator with sufficient bandwidth. The associated collinear phasematching function is ``vertical'' when plotted as a function of the e- and o-ray frequencies: it is broadband in o-ray frequency and has an e-ray bandwidth that is proportional to the inverse of the crystal length. When multiplied by a broadband pump function, the phasematching function for a crystal a few millimetres in length dominates the spectral structure of the photon pairs, with the result that they are uncorrelated in frequency. This situation is illustrated in figure \ref{fig:kdp_jsi} for a 5\,mm long KDP crystal and a broadband frequency-doubled Ti:Sapphire laser centered at 415\,nm. The resulting joint amplitude distribution is clearly highly factorable with a Schmidt number of only $K = 1.01$. Therefore over 99\% of the emission is into the first Schmidt mode.

There is an intuitive physical explanation for how this single-mode emission takes place. As any o-ray downconversion would travel alongside the e-ray pump at their mutual group velocity of $1.97 \times 10^8$\,m/s, its possible emission time is therefore constrained to be contiguous with the highly temporally localized pump pulse. Hence o-ray daughter photons must be emitted into a mode that is similar to the broadband single temporal mode of the pump, and, as the modal structure of the o-ray photon is restricted to a single mode, so is that of the e-ray daughter photon. However, the e-ray downconverted photons travel at a group velocity of $2.02 \times 10^8$\,m/s. An e-ray photon created at the entrance face of the long crystal would have walked off ahead of the pump, whereas one created at the exit face would not, so the temporal mode structure of the e-ray photon is much less localized. Therefore emission is into a single broadband mode for the o-ray photon and a narrower bandwidth single mode for the e-ray photon. Furthermore, the single temporal mode structure of the photon pairs results in exceptionally low jitter in the arrival time of the single photons. This is the starting point for pure heralded photon generation in KDP.

\subsection{Experimental considerations}

The original proposal for factorable state generation in KDP outlined above was based on the plane-wave model of downconversion --- the crystal would be pumped with a collimated beam and only the collinear component of the output was considered. This situation can to a good approximation be modelled entirely analytically. However, to generate photons that can be collected into single-mode fibre one must focus the pump beam into the nonlinear crystal \cite{Dragan2004Efficient-fiber-coupling, Ljunggren2005Optimal-focusing-for-maximal}. Hence any experiment to generate photons in well-defined spatial modes is necessarily different from the plane-wave case and cannot be modelled as such. There are three areas that must be considered in an accurate model: the intrinsic pump focusing itself, collection of photon pairs over a finite range of angles, and any inhomogeneities in the pump beam.

The principal effect of focusing the pump upon the joint spectrum is that, within the pump beam that was previously considered to be collimated, there now exists an angular distribution of wavevectors. This angular spread can be estimated simply through the geometry of the pumping system: the FWHM distribution of propagation angles in the KDP crystal will, in the paraxial limit, be equal to the ratio of the FWHM beam diameter of the collimated pump beam incident on the the preceding lens to the distance from the lens to the crystal (i.e.\ the focal length of the lens). Every angle within this pump distribution will experience different phasematching conditions due to its unique propagation direction relative to the optic axis of the crystal. Hence over the corresponding range of angles exiting the crystal there will be a spread of downconversion wavelengths, each associated with the particular pump angle that gave rise to the pair. The wavelength is highly sensitive to angle; for degenerate collinear phasematching a change in angle of only 1$^{\circ}$ will move the central collinear wavelength by over 10\,nm \cite{Wasylczyk2007A-short-perspective-on-long}.

For phasematching at a fixed central pump wavelength, the effect that the pump focusing has on the total joint spectrum (summed over all pump angles) is more pronounced for the e-ray than the o-ray due to the orientation of the phasematching function. Angular changes in the phasematching conditions appear primarily as a translation of the centre of the phasematching function along the e-ray frequency axis. Summing this distribution over the range of pump angles present broadens the overall e-ray spectral distribution directly. The o-ray is affected only through the energy conservation of the pump envelope function --- as the e-ray wavelength gets shorter, so the o-ray must get longer and vice versa. Although the proportional effect on the o-ray spectrum is much smaller as it is broader to begin with, it is this inverse relationship that gives pump focusing the potential to bring correlations to a state that for collinear pumping would have none.

A related effect is from the fibre coupling collection angle. Pairs are emitted over a finite angular range and collected by coupling optics into single-mode fibre. Due to the angular dependence of the e-ray refractive index, the phasematching conditions vary across this distribution, and can therefore also introduce spectral correlations. To accurately model these two effects, one must take account of both collinear and noncollinear contributions to the two-photon state.

The last factor that one has to consider arises from the method of generating the pump pulses at 415\,nm. In order to obtain high efficiency in the second harmonic generation (SHG) of the Ti:Sapphire laser, the fundamental beam needs to be focused tightly into the nonlinear medium. Hence a broad range of phasematching angles is present in the SHG crystal also, and the differing phasematching conditions across the beam correlates the frequency of the upconverted pulses with their angle of propagation. This angle is mapped onto transverse beam position by the collimating lens following the SHG crystal and the frequency-doubled pulses, destined to pump the downconversion, are therefore spatially chirped --- their central wavelength changes with position across the beam. For a type-I phasematched negative uniaxial SHG crystal, this spatial chirp appears in the principal plane (in which the output is also polarised) as it is only the change in angle in the principal plane that makes a significant difference to the phasematching conditions. In the perpendicular direction the central frequency is independent of position.

\begin{figure}
\begin{center}
\includegraphics[width= 0.7 \textwidth]{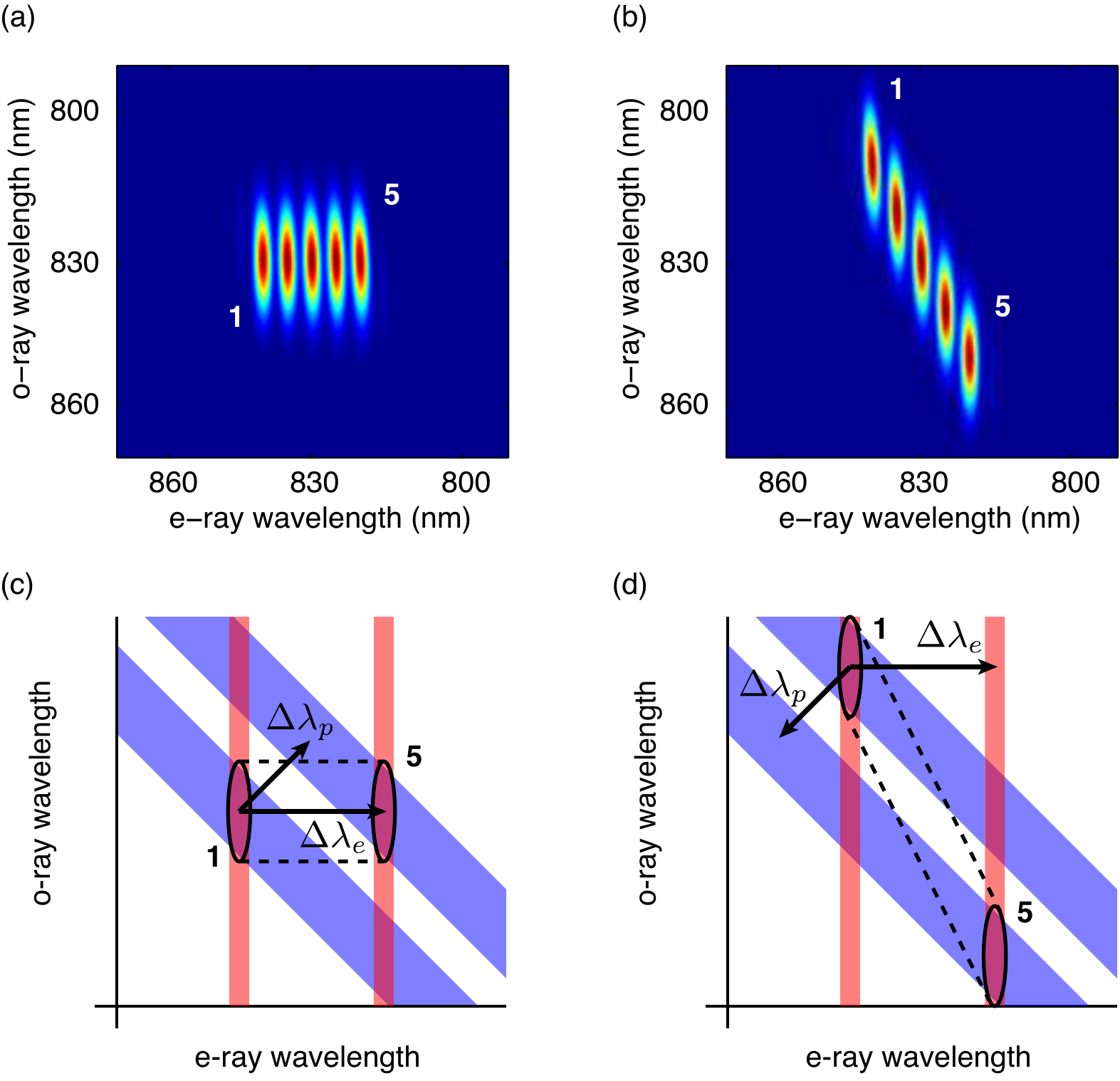}
\caption[Illustration of how pump spatial chirp can affect the joint spectrum in the case of a focused pump in KDP]{An illustration of how pump spatial chirp can affect the joint spectrum in the case of a focused pump in KDP. Five discrete directions of propagation within a continuous pump distribution are considered; (a) and (c) show positive correlation of pump frequency with phasematching angle, (b) and (d) show negative correlation. (a) and (b) contain five discrete joint spectra resulting from the multiplication of five collinear phasematching functions at different phasematching angles by five pump functions whose central wavelength depends on the angle of propagation in the crystal, labelled 1 to 5. Angles change in half degree steps, centered on collinear phasematching at 830\,nm, while pump wavelengths are separated by 5\,nm. (c) and (d) are schematic representations of the difference between the two cases above, with the phasematching functions in red and the pump functions in blue. Both phasematching functions move by $\Delta \lambda_e$; both pump functions move by $\Delta \lambda_p$ but in opposite directions for positive and negative spatial chirp. For positive chirp, the change in the phasematching function is compensated by the corresponding change in the pump wavelength, whereas for negative chirp the phasematching and pump movements compound one another. For a continuous pump distribution, one would sum over the range of the five joint spectra shown, hence resulting in a factorable distribution in the case of positive spatial chirp but a correlated state for negative spatial chirp.}
\label{fig:kdp_pump_chirp}
\end{center}
\end{figure}

When focused into the KDP crystal, position in this spatially chirped beam is once again mapped back onto angle. Therefore each of the different phasematching angles present in the pump distribution in the downconversion crystal will have a different central pump wavelength, hence the effective pump envelope function will be angularly dependent. Graphically, this corresponds to translating the pump function along the line $\omega_e = \omega_o$. When combined with the changes in phasematching function as a result of focusing, this angular variation in the pump function can introduce much stronger correlations into the joint spectral distribution than simply the focusing alone. The effects of a spatially chirped pump beam are demonstrated in figure \ref{fig:kdp_pump_chirp} by plotting a series of five joint spectra corresponding to five collinear phasematching functions at different angles. Each phasematching function is multiplied by a pump function centered at a different wavelength to simulate a spatially chirped beam. It is clear that, in the case of positive correlation between pump frequency and phasematching angle, one could sum over the joint spectra in the range displayed and end up with a total joint spectral intensity uncorrelated in e- and o-ray frequency, whereas in the negatively correlated situation, summing over the joint spectra shown would give a highly correlated state. This situation becomes more complicated when the e-ray output angle is also varied.

\section{Modelling the two-photon joint spectral distribution}
\label{sec:modelling}

To predict the form of PDC emission in a realistic experiment as described above, an accurate method of calculating the effects of the angular distribution of wavevectors present in the pump and downconverted fields upon the joint spectral distribution is required. As contributions from noncollinear wavevectors in the pump and downconverted angular distributions are present in both transverse dimensions, at first glance it would appear that a three dimensional model is essential. However, an accurate model can be constructed by taking into account only two spatial coordinates --- the longitudinal $z$-axis and the transverse axis in the principal plane. As all the angles considered are small and the downconversion takes place in a uniaxial crystal, the angle of propagation out of the principal plane does not cause a significant change in the e-ray refractive index. This is set primarily by the projection of the propagation angle onto the principal plane and hence the problem can be reduced to only two dimensions. Furthermore, as the spatial chirp is in the same plane as the polarisation of the second harmonic beam and this is also the same as the principal plane of the PDC crystal (as the pump is e-polarised), the spatial chirp can also be incorporated into this two-dimensional model. The method presented here can be used for predicting joint spectra from any downconversion process taking place in a uniaxial crystal where the small angles approximation is valid.

\begin{figure*}
\begin{center}
\includegraphics[width= 0.7 \textwidth]{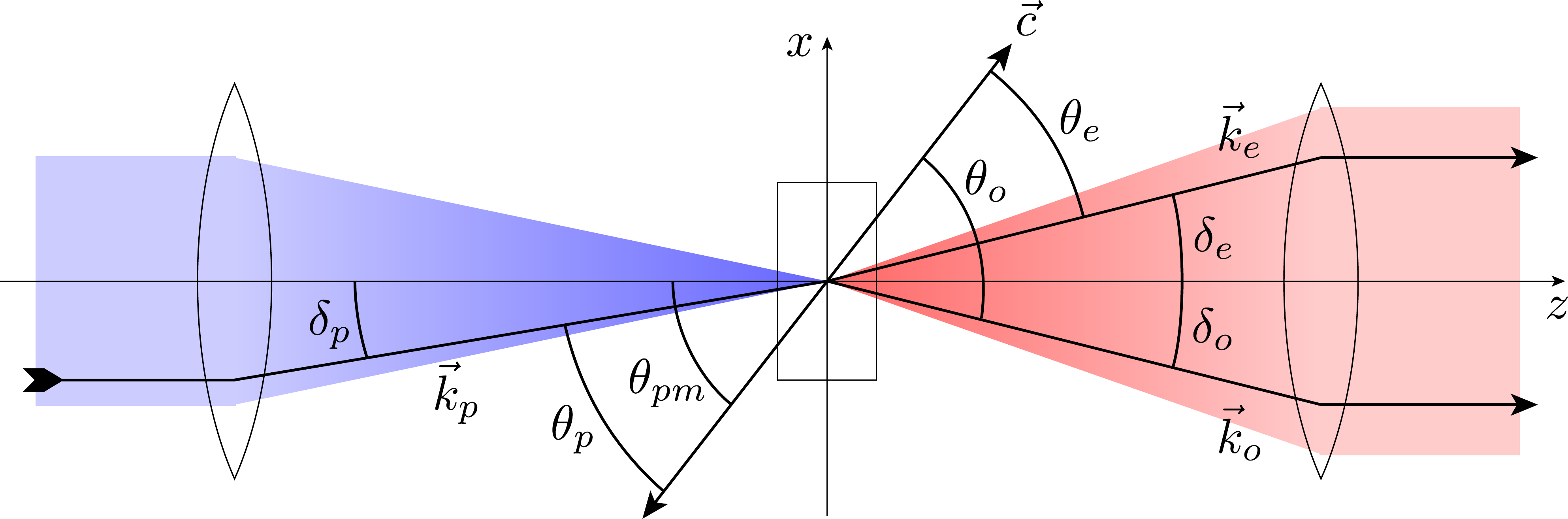}
\caption[Definition of angles in the numerical model]{Definition of angles in the numerical model.}
\label{fig:model_angles}
\end{center}
\end{figure*}

We resolve the pump and downconverted momenta into components in the $x$- and $z$-directions, with the $z$-axis lying along the direction of perfect collinear phasematching and making an angle of $\theta_{pm}$ with the optic axis of the crystal. The optic axis, and hence the principal plane, is set in the $xz$-plane and all the $y$-components of the wavevectors are set to zero. This is illustrated in figure \ref{fig:model_angles}. The wavevector mismatches in the $x$ and $z$ directions are
\begin{align}
\label{eq:model_1}
\Delta k_x & = k_{e,x}(\omega_p, \delta_p) - k_{e,x}(\omega_e, \delta_e) - k_{o,x}(\omega_o) \nonumber \\
\Delta k_z & = k_{e,z}(\omega_p, \delta_p) - k_{e,z}(\omega_e, \delta_e) - k_{o,z}(\omega_o).
\end{align}
The components of the pump wavevector are simply
\begin{align}
\label{eq:model_1b}
k_{e,x}(\omega_p, \delta_p) & = k_{e}(\omega_p, \theta_p) \sin{\delta_p}, \nonumber \\
k_{e,z}(\omega_p, \delta_p) & = k_{e}(\omega_p, \theta_p) \cos{\delta_p},
\end{align}
where $k_{\mu}(\omega_{\mu}, \theta_{\mu}) = |\vec{k}_{\mu}(\omega_{\mu}, \theta_{\mu})|$ and $\theta_{\mu}$ is the angle subtended by each ray and the optic axis, $\theta_{\mu} = \theta_{pm} - \delta_{\mu}$. Similar expressions can be written down for the e- and o-ray components. By imposing perfect transverse phasematching so that $\Delta k_x = 0$ we find
\begin{equation}
\label{eq:model_3}
\delta_o = \arcsin{\frac{k_{e,x}(\omega_p, \delta_p) - k_{e,x}(\omega_e, \delta_e)}{k_o(\omega_o)}}.
\end{equation}
So for a given set of \{$\omega_p$, $\delta_p$, $\delta_e$\} the angle $\delta_o$ that gives perfect transverse phasematching at every pair of e- and o-ray frequencies can be found.

From (\ref{eq:model_1}) and (\ref{eq:model_3}) we see that the longitudinal wavevector mismatch is
\begin{multline}
\label{eq:model_4}
\Delta k_z(\omega_e, \omega_o, \delta_p, \delta_e) = k_e(\omega_e + \omega_o, \theta_p) \cos{\delta_p} - k_e(\omega_e, \theta_e) \cos{\delta_e} \\
- k_o(\omega_o) \cos{\left[ \arcsin \left(  \frac{k_{e,x}(\omega_e + \omega_o, \delta_p) - k_{e,x}(\omega_e, \delta_e)}{k_o(\omega_o)} \right) \right]},
\end{multline}
where the $x$-components are as defined in (\ref{eq:model_1b}). Therefore, for set values of the pump angle and e-ray collection angle, from this relationship can be found the wavevector mismatch for each pair of frequencies of the daughter photons. By substituting $\Delta k_z(\omega_e, \omega_o, \delta_p, \delta_e)$ into the longitudinal phasematching condition
\begin{equation}
\label{eq:model_5}
\phi(\omega_e, \omega_o, \delta_p, \delta_e) = e^{\frac{i \Delta k_{z} L}{2}} \mathrm{sinc} \left( \frac{\Delta k_{z} L}{2} \right)
\end{equation}
the joint spectrum can be plotted for any pair of angles $\delta_p$ and $\delta_e$.

In order to calculate the joint spectrum of a realistic photon pair, however, a set of these plane wave solutions must be taken over the full range of pump and collection angles. Firstly, the focused pump is represented by a superposition of plane waves, summed over the angle $\delta_p$:
\begin{equation}
\label{eq:model_6}
E_{p}^{(+)}(\vec{r},t) = A_{p} \int_{0}^{\infty} d\omega_{p} \int_{-\frac{\pi}{2}}^{\frac{\pi}{2}} d\delta_{p} \alpha(\omega_{p}, \delta_p) \exp{\left[ i \left( \vec{k}_{e}(\omega_{p}, \delta_{p}).\vec{r} - \omega_{p} t \right) \right]}.
\end{equation}
The Gaussian angular dependence of the pump is written implicitly in $\alpha(\omega_p, \delta_p)$, and for linear spatial chirp the central pump frequency becomes a function of the same angle, $\delta_p$:
\begin{align}
\label{eq:model_7}
\alpha(\omega_p, \delta_p) & = \alpha(\omega_e + \omega_o, \delta_p) \\
& = \exp{\left[ - \left( \frac{\omega_e + \omega_o - 2(\omega_0 + q \delta_p)}{\sigma} \right)^2 \right]} \exp{\left[ - \left(\frac{\delta_p}{\sigma_L}\right)^2 \right]}, \nonumber
\end{align}
where $q$ is a constant, defined at $\omega_0$, describing the rapidity with which the central pump frequency changes across the angular distribution of the pump. The angular bandwidth of the pump, set by the strength of the lens before the downconversion crystal, is given by $\sigma_L$ and the central wavevector of the pump distribution lies along the $z$-axis. Secondly, the collection of the downconversion into single-mode fibres is modelled by two Gaussian angular filter functions, one dependent on $\delta_e$ and the other on $\delta_o$:
\begin{equation}
\label{eq:model_8}
g_F(\delta_e) = \sqrt[4]{\frac{2}{\pi \sigma_{F}^2}} \exp{\left[ - \left( \frac{\delta_e}{\sigma_F} \right)^2 \right]}
\end{equation}
and similarly for $\delta_o$, where it is assumed that the angular acceptance bandwidth $\sigma_F$ is identical for both photons and the peak transmission is along the $z$-axis (at $\delta_e = \delta_o = 0$).

Incorporating these relationships into the expression for the two-photon state, the total joint spectral amplitude at the fibre output becomes
\begin{equation}
\label{eq:model_9}
f(\omega_{e}, \omega_{o}) = \int_{-\frac{\pi}{2}}^{\frac{\pi}{2}} d\delta_{p} \int_{-\frac{\pi}{2}}^{\frac{\pi}{2}} d\delta_{e} \int_{-\frac{\pi}{2}}^{\frac{\pi}{2}} d\delta_{o} \alpha(\omega_{e} + \omega_{o}, \delta_p) \phi(\omega_{e}, \omega_{o}, \delta_p, \delta_e) g_F(\delta_e) g_F(\delta_o)
\end{equation}
For small angles of emission, $\delta_o \approx - \delta_e$ and $g_F(\delta_e) g_F(\delta_o) = (g_F(\delta_e))^2$; this effectively assumes that if the e-ray photon passes the filter and is collected into the fibre, the o-ray photon is also. Making this approximation reduces by one the number of sums in the numerical calculation hence reducing the processing time required. Equation (\ref{eq:model_9}) reduces to
\begin{equation}
\label{eq:model_10}
f(\omega_{e}, \omega_{o}) = \int_{-\frac{\pi}{2}}^{\frac{\pi}{2}} d\delta_{p} \int_{-\frac{\pi}{2}}^{\frac{\pi}{2}} d\delta_{e} \alpha(\omega_{e} + \omega_{o}, \delta_p) \phi(\omega_{e}, \omega_{o}, \delta_p, \delta_e) g_F^2(\delta_e)
\end{equation}
where the Gaussian filter function $g_F^2(\delta_e)$ is the angular distribution of the \textit{pairs} that are collected into the fibre. Equation (\ref{eq:model_10}) forms the basis upon which the two-photon state generated by a focused pump and collected into single-mode fibre can be modelled numerically.

A numerical simulation based on the equations in this section was implemented in Matlab. The program was run to generate joint spectra over a range of values of parameters such as the pump angular distribution, collected mode angle, crystal length, and the central phasematching angle (the crystal angle, $\theta_{pm}$). This allowed the effects of different experimental configurations upon the properties of the joint spectrum, in particular its factorability, to be determined. The factorability in each configuration was quantified using the Schmidt number, $K$, found from the SVD of the final joint spectral amplitude summed over the relevant angles.

The large number of variables in the model make it time-consuming to search over all of them for an optimal solution. However, several can be fixed, or at least have their ranges reduced, by taking into account the experimental constraints. For example, the \textit{magnitude} of the spatial chirp on the pump beam is fixed by the type of crystal used for SHG, its length, and how tightly the pump is focused into it --- all dictated by the need for efficient frequency conversion. This also sets the pump bandwidth. On the other hand, the \textit{direction} of the spatial chirp can be switched simply by rotating the downconversion crystal by 180$^{\circ}$ about the central pump direction (the $z$-direction) to swap the sense of the pump frequency shift with angle in the principal plane of the KDP crystal. For the purposes of this discussion, positive spatial chirp is defined as pump wavelength increasing with $\delta_p$ (so $q$ is in fact negative), or in other words, pump frequency dropping as the angle between the pump and the optic axis is reduced (see figure \ref{fig:model_angles}).

However, this has only eliminated two parameters from our search. Other source parameters that are relatively simple to change, and hence must be varied in the model, include the pump focusing, collection angle, crystal angle, and crystal length (so long as a range of crystals are available for use). Yet, the range of values that these can take is limited by the conditions required for factorability and other experimental limitations.

The pump focusing conditions on one hand should be set to get as close to a Fresnel number of one as possible to maximize generation in the single spatial mode that gets coupled into the fibres \cite{Ljunggren2005Optimal-focusing-for-maximal}, but on the other hand, due to the effect of spatial walkoff in the crystal, the spot size cannot be made too small. The pair collection angle should be such that the fibre coupling optics map the pumped volume of the crystal into the core of the fibre, but cannot be too large because, as described earlier in this section, increasing the collection angle can introduce additional correlations to the joint spectrum. The crystal length must be sufficiently large relative to the inverse pump bandwidth that the width of the crystal phasematching function is significantly smaller than the pump bandwidth \cite{URen2005Generation-of-pure-state-single-photon}, but not so long that spatial walkoff becomes a problem. We require our photons to be approximately degenerate around 830\,nm; hence $\theta_{pm}$ must be restricted to be close to the degenerate collinear phasematching angle at 830\,nm ($\theta_{pm} = 67.8^{\circ}$).

\begin{figure}
\begin{center}
\includegraphics[width= 0.8 \textwidth]{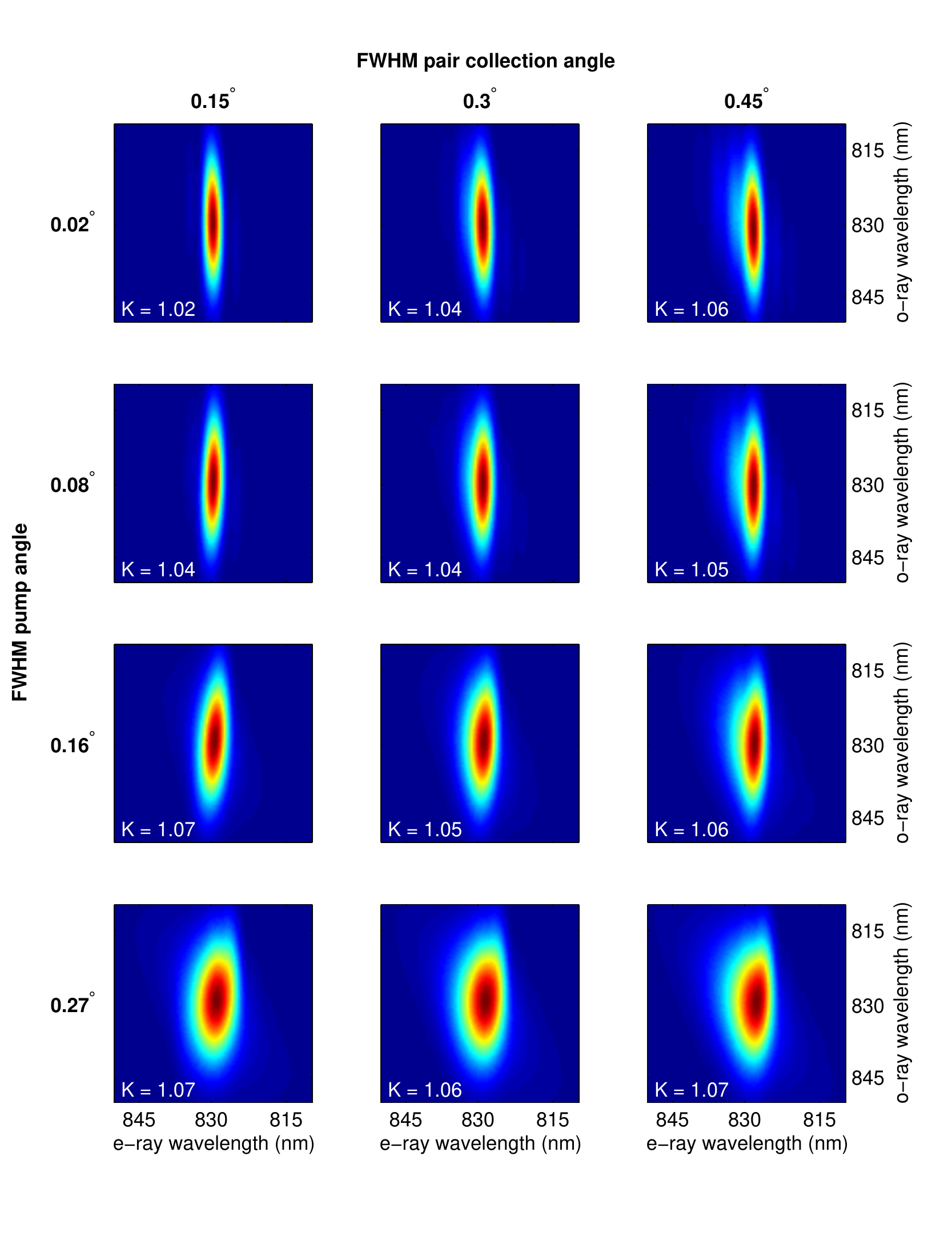}
\caption[Predicted joint spectral intensities for a range of pump and collection angles for positive spatial chirp]{Predicted joint spectral intensities for a range of pump and collection angles for positive spatial chirp. See text for details.}
\label{fig:model_plots_positive}
\end{center}
\end{figure}

\begin{figure}
\begin{center}
\includegraphics[width= 0.8 \textwidth]{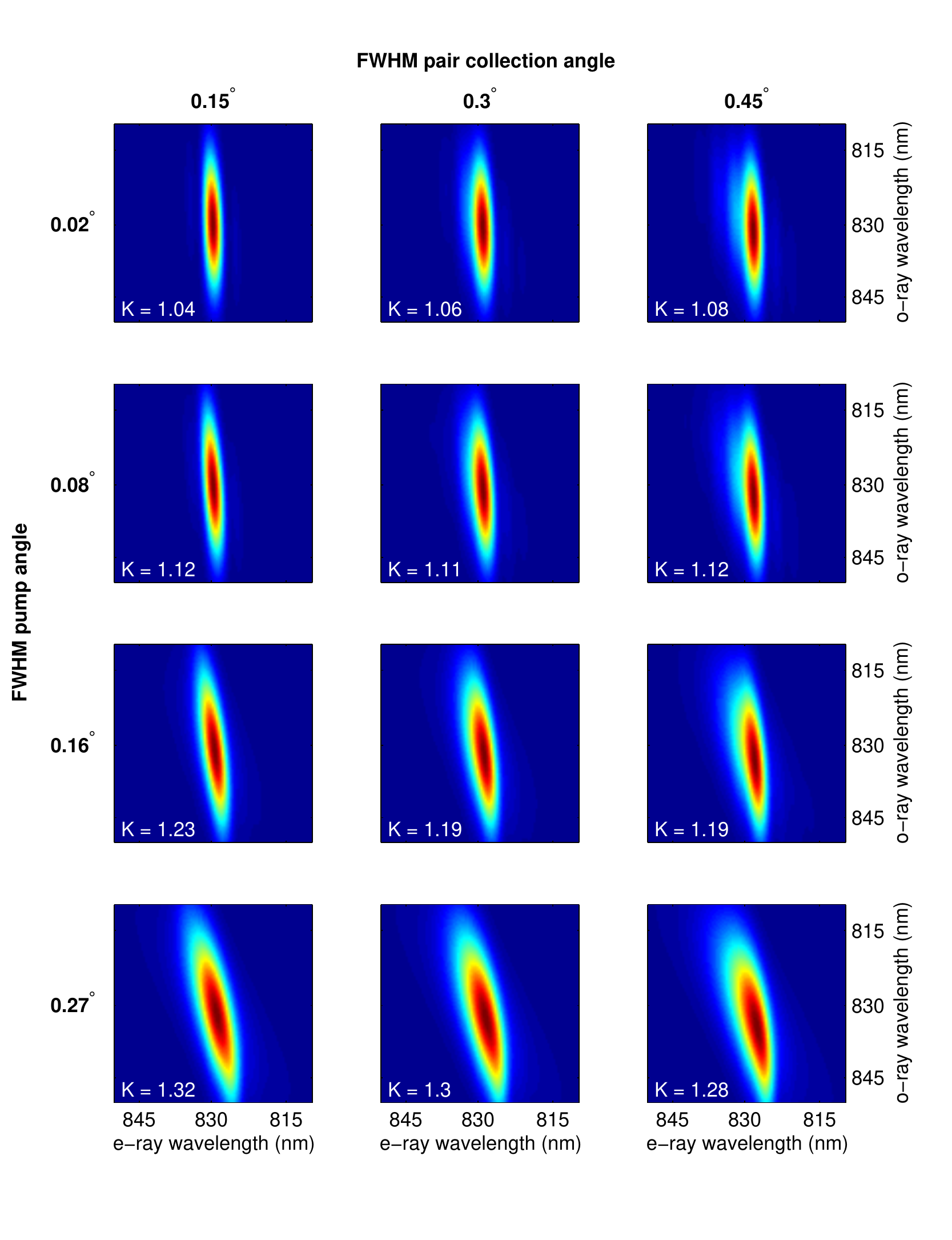}
\caption[Predicted joint spectral intensities for a range of pump and collection angles for negative spatial chirp]{Predicted joint intensities for a range of pump and collection angles for negative spatial chirp. See text for details.}
\label{fig:model_plots_negative}
\end{center}
\end{figure}

Figures \ref{fig:model_plots_positive} and \ref{fig:model_plots_negative} demonstrate some of the results of the model for positive and negative spatial chirps respectively. The joint spectra were all evaluated on a 100$\times$100 grid in frequency space over a range of 40\,nm for both the e- and o-rays. These plots are for a 5\,mm crystal and 7.5\,nm of spatial chirp across the 700\,$\mu$m FWHM diameter of the pump intensity distribution before the pump lens. The intensity FWHM pair collection angle is 0.15$^{\circ}$, 0.3$^{\circ}$, and 0.45$^{\circ}$ left to right while the FWHM angular intensity of the pump distribution is 0.02$^{\circ}$, 0.08$^{\circ}$, 0.16$^{\circ}$, and 0.27$^{\circ}$ top to bottom, corresponding to focal lengths of 2000\,mm, 500\,mm, 250\,mm, and 150\,mm respectively for this pump beam diameter. The sums were performed over all combinations of 11 equally spaced angles in both the pump and collected distributions. It can be seen that the behaviour of the joint spectrum is very different for opposite spatial chirps. However, generally the distributions become broader and more correlated as the level of focusing is increased. As expected, spatial chirp does not affect the two-photon state generated by light focusing as all the frequency components in the pump beam experience approximately the same phasematching conditions.

\begin{figure}
\begin{center}
\includegraphics[width= 0.5 \textwidth]{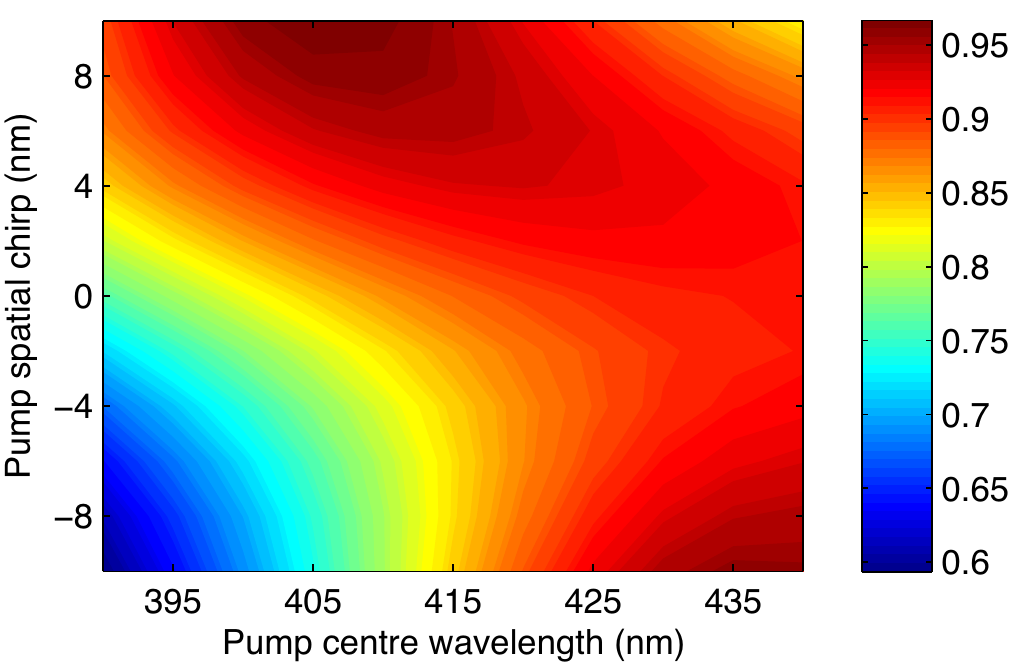}
\caption{Numerical optimization plot showing the expected purity as a function of pump centre wavelength against spatial chirp. In The FWHM pump angle was 0.16$^{\circ}$ and the pair collection angle was 0.30$^{\circ}$.}
\label{fig:model_opt_pumpwl_vs_chirp}
\end{center}
\end{figure}

By running the model repeatedly it was possible to generate plots of the expected purity to show the optimum values of various parameters for factorable state generation; one is shown in figure \ref{fig:model_opt_pumpwl_vs_chirp}. As noted above, it was difficult to control the magnitude of the spatial chirp experimentally, though interestingly a positive spatial chirp with our experimentally observed value (7.5\,nm across the FWHM beam diameter) yields almost the highest purity for a pump beam centered at 415\,nm given our other experimental parameters.

The clearest result from this model is that, in order to prepare a state that is anywhere close to factorable with high count rates, it is essential to understand the effects of spatial chirp. Although it was not considered in the initial proposals for factorable state generation, some spatial chirp can in fact be a help rather than a hinderance on the road to factorability, as shown in figure \ref{fig:model_opt_pumpwl_vs_chirp}. However, the spatial chirp and the crystal must be correctly orientated relative to one another --- if they are not, the resulting state will be correlated. If the spatial chirp, pump focusing, and crystal are carefully matched though, the joint state can be made highly factorable. This can be thought of as the change in pump frequency with angle due to the spatial chirp offsetting the shift in central downconversion wavelength that occurs as a result of the different phasematching conditions across the beam due to focusing. Hence these two effects can be made to cancel one another out and produce a factorable state. The exact focusing and crystal parameters required to achieve this are most easily found through the numerical model presented here as no simple relationship exists.

\section{Measurement of joint spectral intensity}

The final source configuration decided upon to yield the best combination of factorability and pair generation rate was a 5\,mm KDP crystal cut for type-II phasematching at 830\,nm, with the pump focused by a 250\,mm focal length lens, and the emission from the crystal collimated by a 150\,mm lens placed one focal length afterwards. This gave a FWHM pump angular intensity distribution of 0.16$^{\circ}$ and an intensity FWHM pair collection angle of 0.30$^{\circ}$. A more detailed description of the source setup is given in section \ref{sec:homi}.

\begin{figure}
\begin{center}
\includegraphics[width= 0.8 \textwidth]{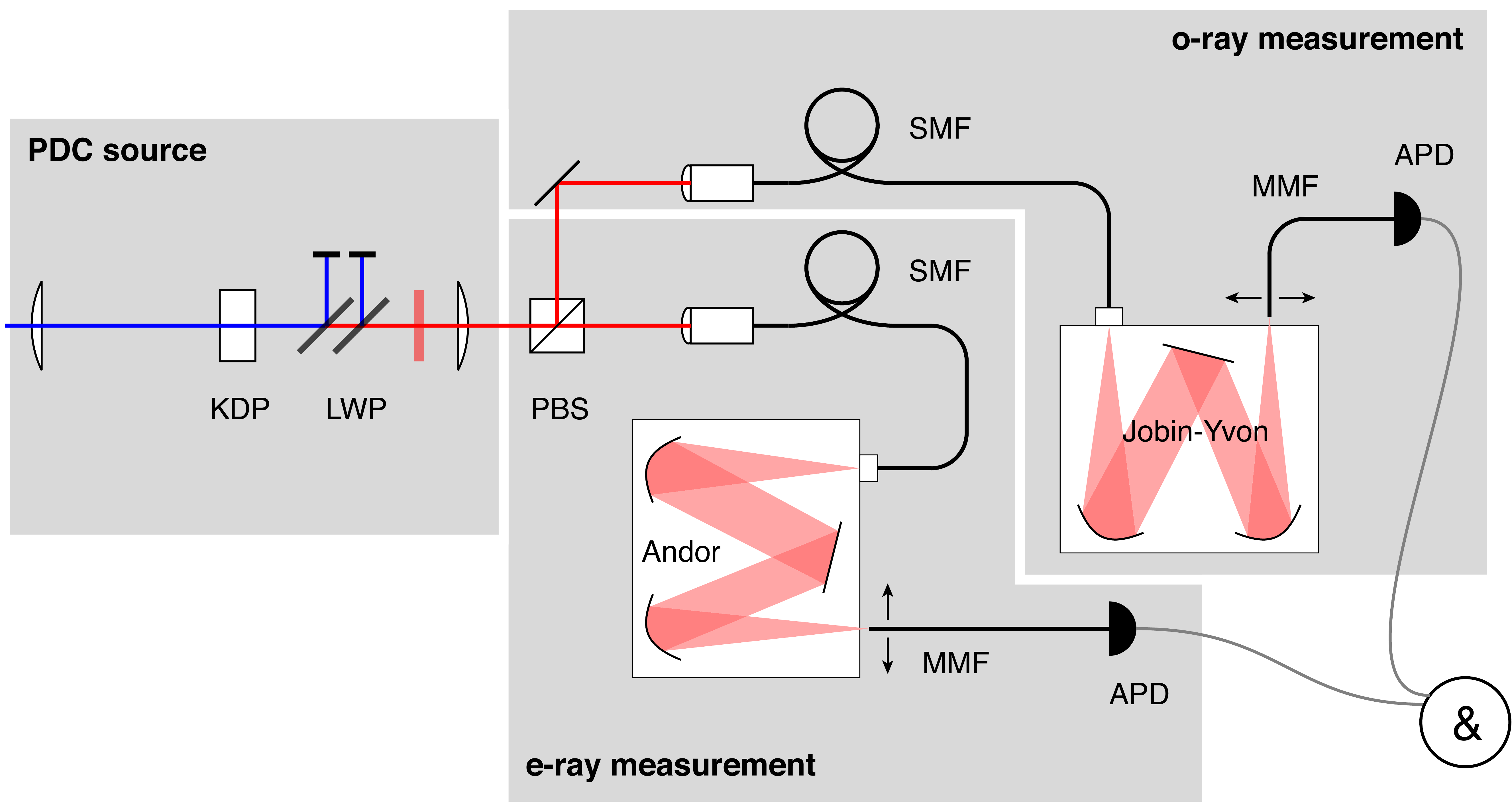}
\caption[Apparatus for measurement of joint spectra]{Apparatus for measurement of joint spectra. KDP = downconversion crystal, LWP = long-wave-pass filter, PBS = polarising beamsplitter, SMF = single-mode fibre, MMF = multimode fibre, APD = avalanche photodiode.}
\label{fig:jsi_app}
\end{center}
\end{figure}

The numerical model outlined above suggested that these parameters would yield highly factorable states for the correct direction of spatial chirp; the calculated amplitude distributions had Schmidt numbers of $K = 1.05$  for positive spatial chirp but $K = 1.19$ if the spatial chirp was negative. In order to test the results of the model, we made a direct measurement of the joint spectrum of the photon pairs using two monochromators \cite{Kim2005Measurement-of-the-spectral-properties} for both directions of spatial chirp, as described in reference \cite{Mosley2008Focusing-on-factorability:-Space-time} and shown  in figure \ref{fig:jsi_app}. This allowed the quantification of the degree of spectral correlation between the downconverted pairs, although it provided no information about the phase of the joint spectral amplitude and therefore the degree of temporal correlation remained unknown.

\begin{figure}
\begin{center}
\includegraphics[width= 0.65 \textwidth]{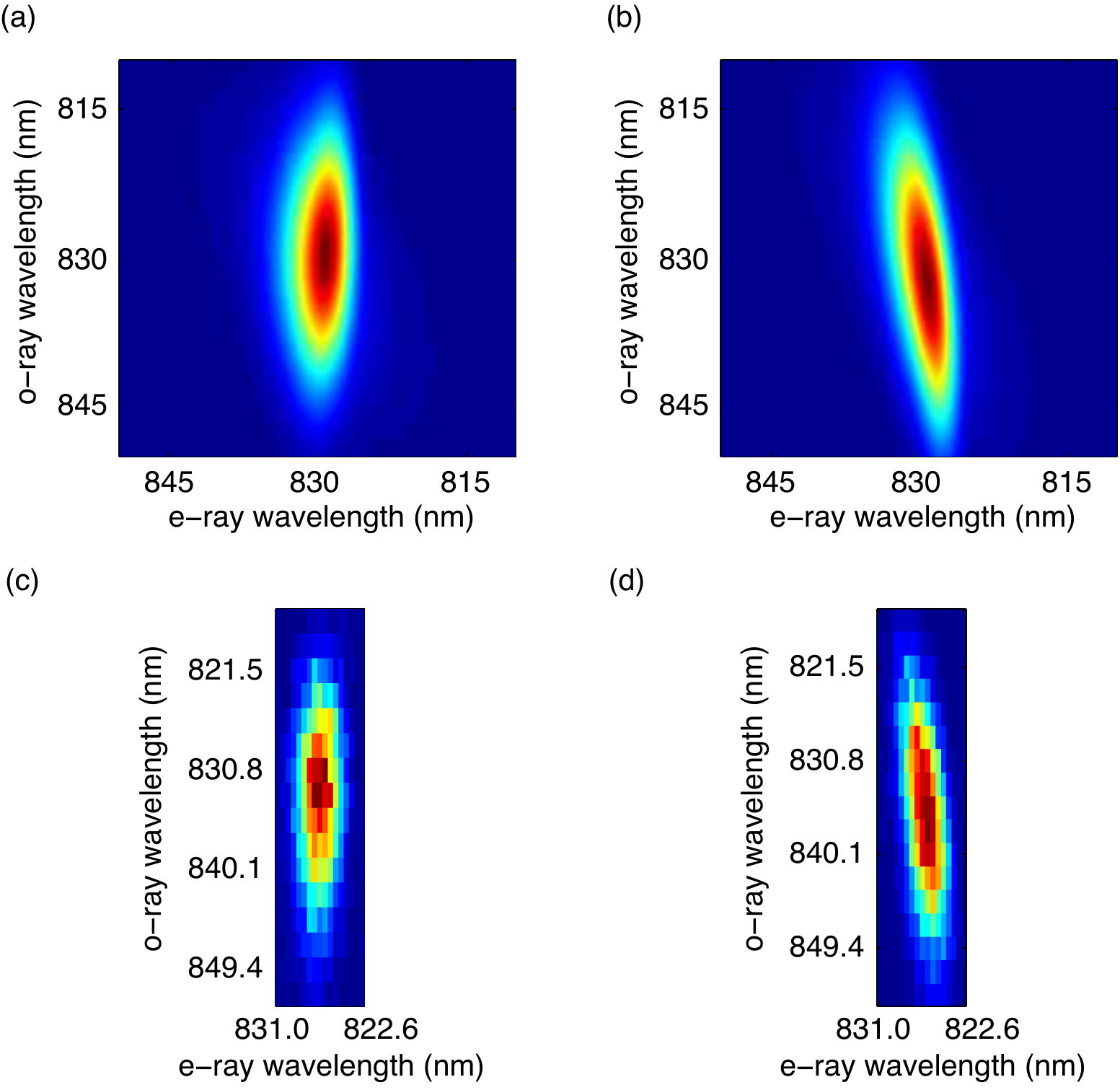}
\caption[Numerically calculated joint spectra alongside measured joint probability distributions]{Joint spectra calculated from the numerical model for the measured experimental parameters with positive spatial chirp (a) and negative spatial chirp (b). Measured joint spectral probability distributions for positive spatial chirp (c) and negative spatial chirp (d).}
\label{fig:jsi_data_model}
\end{center}
\end{figure}

The experimental data are presented in figure \ref{fig:jsi_data_model} alongside the joint spectral distributions predicted by the numerical model given the same source parameters. It can be seen that the agreement between the two is very good for both positive and negative spatial chirp. The experimental spectral intensity distribution is highly factorable in the case of positive chirp: if flat spectral phase is assumed across the corresponding joint amplitude distribution, the associated Schmidt number is 1.02. On the other hand, for negative chirp the joint state is much less factorable, with $K = 1.14$.

This can be compared more quantitatively with the output of the model by finding the Schmidt number for the amplitude distribution predicted by the model assuming that we have no information about the phase. This was done by calculating the SVD of the square root of the joint spectral intensity rather than directly from the joint amplitude. The comparison between the model and the data is displayed in table \ref{table:schmidt_comp}, along with the figures for the model inclusive of its phase. It can be seen that the model accurately predicts the Schmidt number expected for the measured intensity distribution for both directions of chirp, but the true Schmidt number for the predicted amplitude distribution including phase is a little higher. It is from this final Schmidt number that the expected visibility for the interference of heralded photons from each distribution can be calculated. For a positively-chirped pump beam, this gives a projected maximum purity and visibility of just over 0.95.

\begin{table}
\renewcommand{\arraystretch}{1.3}
\begin{center}
\small{
\begin{tabular}{p{4cm} p{1.5cm} p{2cm} p{1.5cm} p{1.5cm}}	\hline	
				& \multicolumn{2}{l}{Positive chirp}		& \multicolumn{2}{l}{Negative chirp}		\\ \hline
				& $K$			& $\mathcal{P}$		& $K$		& $\mathcal{P}$		\\	\hline
Data				& 1.02			& 0.979			& 1.12			& 0.894			\\
Model (no phase)	& 1.03			& 0.970			& 1.17			& 0.854			\\
Model (with phase)	& 1.05			& 0.953			& 1.19			& 0.839			\\ \hline
\end{tabular}}
\caption{Comparison of Schmidt number ($K$) and projected purity ($\mathcal{P}$) for the measured joint intensity distributions, the model without any phase, and the model with phase for both spatial chirp directions.}
\label{table:schmidt_comp}
\end{center}
\end{table}

A further method of comparing the correlations present in each spectrum is to plot the e-ray wavelength at which the maximum count rate occurs at every o-ray wavelength. For the experimental data, this is most accurately done by fitting each data slice at constant o-ray wavelength with a Gaussian distribution in e-ray wavelength. The centres of these fits can then be plotted against the o-ray wavelength at which they were taken. As the results from the model are at much higher resolution and are therefore more smoothly varying, it is adequate for these plots to simply take the e-ray wavelength that gives the maximum value of the intensity distribution at each o-ray wavelength.

\begin{figure}
\begin{center}
\includegraphics[width= 0.75 \textwidth]{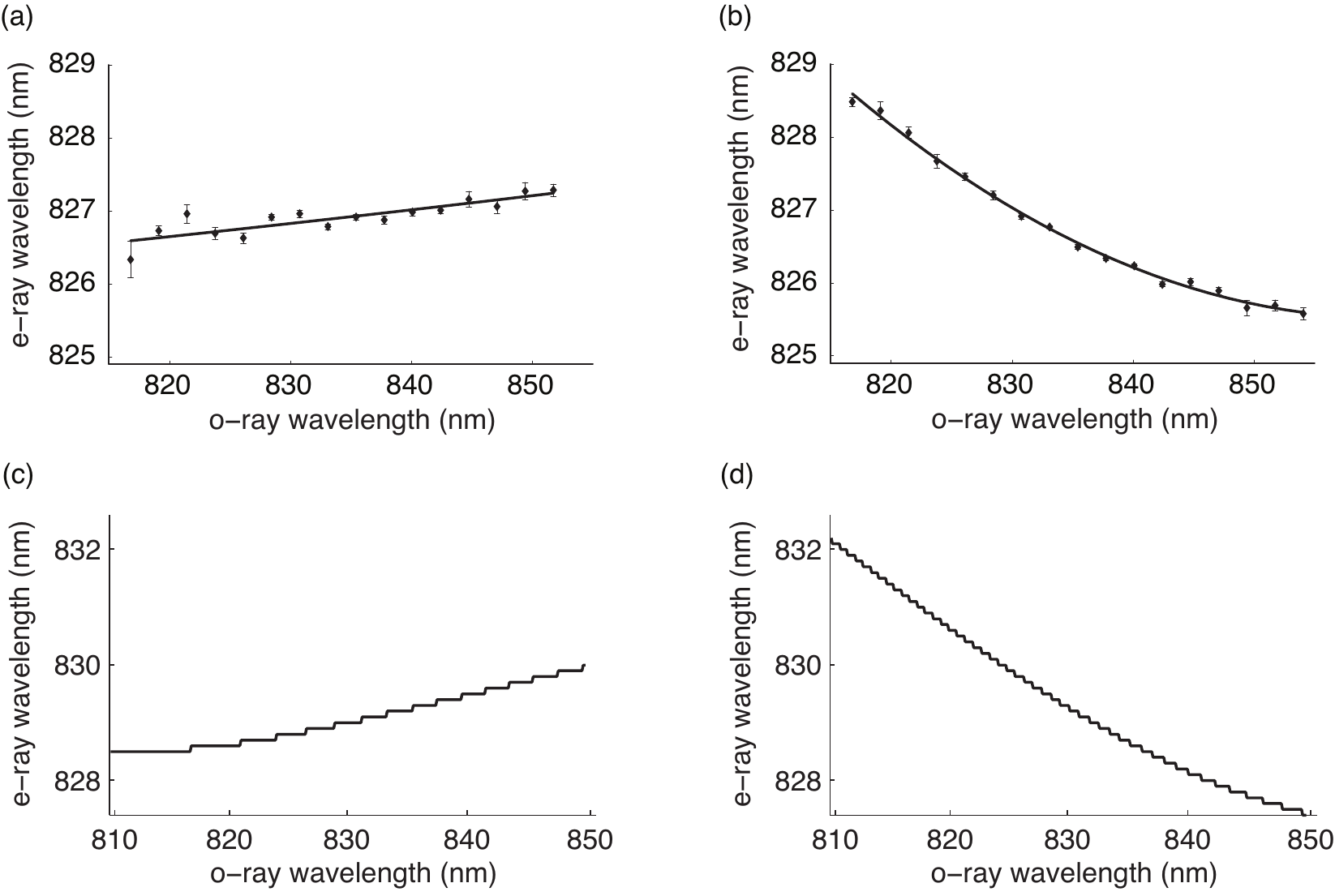}
\caption[Correlation between e- and o-ray photons from measured spectra]{Central e-ray wavelengths of Gaussian fits to slices of constant o-ray wavelength from measured joint spectra for positive (a) and negative (b) spatial chirp. The lines are quadratic fits to the data to guide the eye. (c) and (d) show for positive and negative spatial chirp respectively the e-ray wavelength giving maximum spectral intensity as a function of o-ray wavelength for the calculated joint spectra using the experimental parameters.}
\label{fig:jsi_centres_both}
\end{center}
\end{figure}

The results of this process are shown in figure \ref{fig:jsi_centres_both}. For the measured spectra, in the case of the uncorrelated spectrum resulting from positive spatial chirp on the pump beam, it can be seen that the centre of the e-ray spectrum is almost constant over the entire o-ray spectrum. The fit line to these e-ray central wavelength points changes by only half a nanometre over 30\,nm of o-ray bandwidth, again demonstrating the factorable nature of this state. However, for negative spatial chirp the e-ray centre wavelength is anti-correlated with o-ray wavelength. This can be compared to the results from the model which display the same behaviour.

\begin{figure}
\begin{center}
\includegraphics[width= 0.65 \textwidth]{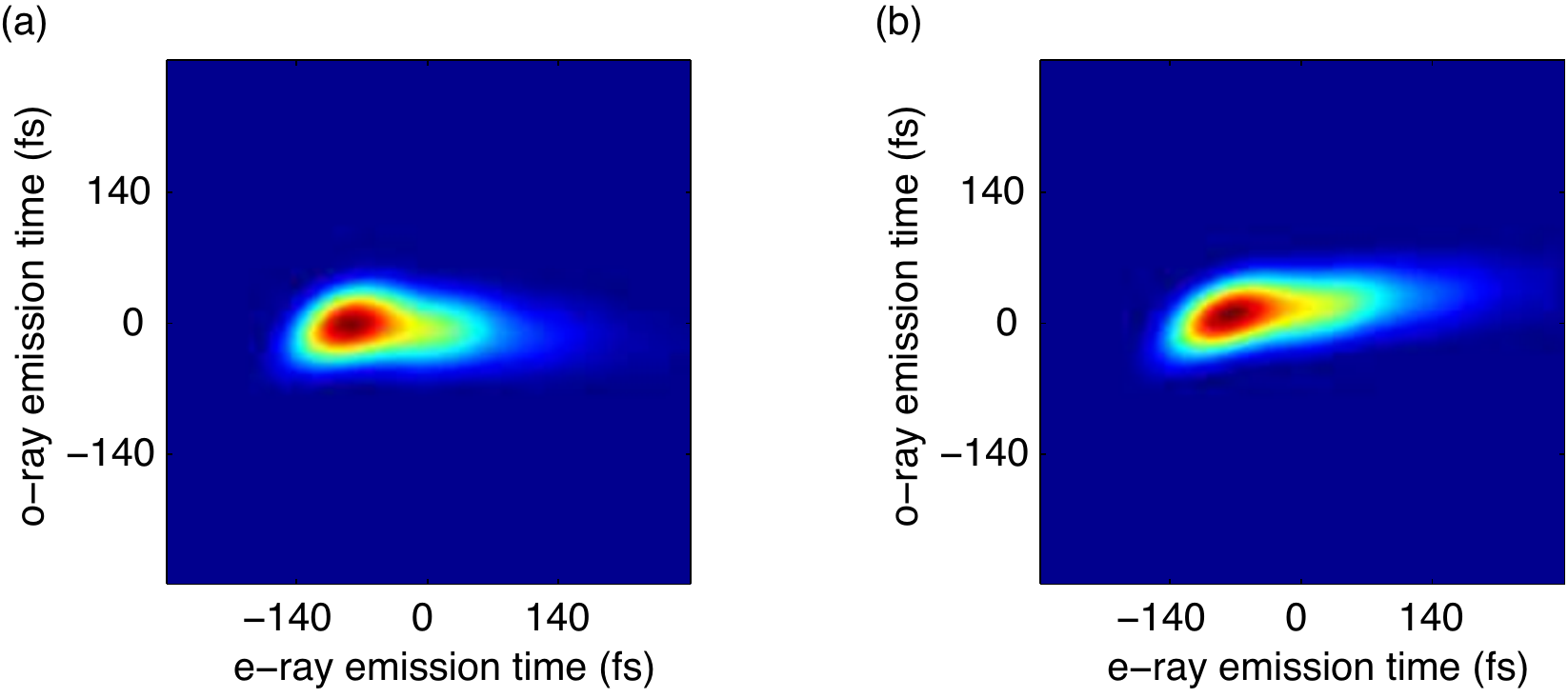}
\caption[Temporal structure of the two-photon state]{The simulated joint temporal intensity distributions for downconversion from KDP in the case of positive (a) and negative (b) spatial chirp. For positive chirp, it can be seen that the temporal structure is approximately single mode. It is this that gives the photons their exceptionally low timing jitter.}
\label{fig:kdp_model_temporal}
\end{center}
\end{figure}

The high level of agreement between the simulated and measured spectral data allows the numerical model to be used to estimate the temporal structure of the photon pairs. This cannot be done from the experimental spectra as no phase information is available. However, by taking the numerical Fourier transform of the simulated spectral amplitudes, we gain some insight into the temporal distribution of the photon pairs. The simulated joint temporal intensities are shown in figure \ref{fig:kdp_model_temporal}; the difference between the lack of any temporal correlation in the case of positive spatial chirp and the tilt of the joint temporal intensity distribution for negative chirp can be clearly seen. Furthermore, the exceptionally short duration of the o-ray photons is evident, along with the precise timing of both photons in the case of positive chirp.

Finally, the measurements on the joint spectrum provide a means of estimating the bandwidths of the daughter photons from the marginal distributions of both the e-ray and o-ray. The marginal frequency distribution of each photon is found by integrating over the frequency of the other photon. These are compared with the estimations of the bandwidths found from the marginal distributions calculated by the numerical model in table \ref{table:bandwidth_comp}. The bandwidths from both agree reasonably well, especially in the trends displayed: the e-ray bandwidths remain approximately constant for both spatial chirps but the o-ray bandwidths are significantly broader in the case of negative chirp. However, the model overestimates the bandwidths in all cases.

\begin{table}
\renewcommand{\arraystretch}{1.3}
\begin{center}
\small{
\begin{tabular}{p{2cm} p{2cm} p{3cm} p{2cm} p{2cm}}		\hline
		& \multicolumn{2}{l}{Positive chirp}		& \multicolumn{2}{l}{Negative chirp}		\\ \hline
		& $\Delta \lambda_e$\,(nm)	& $\Delta \lambda_o$\,(nm)	& $\Delta \lambda_e$\,(nm)	& $\Delta \lambda_o$\,(nm)	\\	\hline
Model	& 5.4				& 21.0		& 5.4				& 25.0		\\
Data		& 3.5				& 16.4		& 3.4				& 19.8		\\	\hline
\end{tabular}}
\caption{Comparison of FWHM bandwidths of the marginal frequency distributions calculated from the model and the measured joint spectral data.}
\label{table:schmidt_comp}
\end{center}
\end{table}

\section{Direct test of photon purity}
\label{sec:homi}

Having demonstrated the veracity of the predictions of our model through measuring experimentally the factorable joint spectral distribution of the emission from our source, it was then necessary to prove that the single photons conditionally prepared from the downconverted pairs were indeed in pure states. This was done by observing Hong--Ou--Mandel interference between the heralded single photons from two independent sources. In order to see high-visibility interference, the photons arriving at the beamsplitter must be both identical and pure; the visibility sets a lower bound on the purity and therefore also determines the factorability of the photon pairs \cite{Grice1997Interference-and-Indistinguishability-in-Ultrafast}. In this section we describe the source setup and alignment techniques and present the experimental results in this context.

It can be shown that upon interfering two distinguishable, mixed photons with density operators $\hat{\rho}_1$ and $\hat{\rho}_2$ at a 50:50 beamsplitter the visibility of the subsequent HOMI dip is given by \cite{Ekert2002Direct-Estimations-of-Linear}:
\begin{equation}
\label{eq:v1}
V = \mathrm{Tr}\left(\hat{\rho}_1 \hat{\rho}_2 \right).
\end{equation}
This non-unit visibility contains contributions from the two photons both individually in their impurity and jointly in their distinguishability. In order to calculate the mean photon purity we wish to separate these two effects, and to do so we consider a measure of the overlap of two states known as the operational distance, defined as \cite{Lee2003Operationally-Invariant-Measure}:
\begin{equation}
\label{eq:v2}
O(\hat{\rho}_1, \hat{\rho}_2) = || \hat{\rho}_1 - \hat{\rho}_2 ||^2
\end{equation}
where $||A||^2 = \mathrm{Tr}\left(A^{\dag} A\right)$ is the Frobenius norm. Expanding this expression we find
\begin{equation}
\label{eq:v3}
O(\hat{\rho}_1, \hat{\rho}_2) = \mathrm{Tr}\left(\hat{\rho}_1^2\right) + \mathrm{Tr}\left(\hat{\rho}_2^2\right) - 2 \mathrm{Tr}\left(\hat{\rho}_1 \hat{\rho}_2\right),
\end{equation}
and rearranging using (\ref{eq:v1}) along with the definition of purity in (\ref{eq:matrix_2}) the visibility is then given by:
\begin{equation}
\label{eq:v4}
V = \frac{\mathcal{P}_1(\hat{\rho}_1) + \mathcal{P}_2(\hat{\rho}_2) - O(\hat{\rho}_1, \hat{\rho}_2)}{2}.
\end{equation}
$O(\hat{\rho}_1, \hat{\rho}_2)$ takes values between zero for $\hat{\rho}_1 = \hat{\rho}_2$ and two for completely distinguishable states. Hence we see that the visibility sets a lower bound on the mean photon purity; for given $V$ the corresponding minimum value of $\left( \mathcal{P}_1+ \mathcal{P}_2 \right)$ occurs where $O(\hat{\rho}_1, \hat{\rho}_2) = 0$ and any deviation away from perfect overlap increases the inferred mean purity.

However, expressing the visibility in the form of (\ref{eq:v4}) only allows one to find a minimum bound on the average purity and not to calculate its value. In fact it can be seen that the effects of mixedness and distinguishability upon the outcome of this measurement cannot in general be separated. Only in two trivial cases, those of either unit purity for both states or perfect overlap, can the mean purity be calculated. Beyond this, the independent measurement of the marginal distributions of $\hat{\rho}_1$ and $\hat{\rho}_2$ yields no additional information as $O(\hat{\rho}_1, \hat{\rho}_2)$ cannot be calculated without prior knowledge of the purities. If one assumes that the two states are pure the operational distance can be found to be:
\begin{equation}
\label{eq:v5}
O_{\mathrm{pure}}(\hat{\rho}_1, \hat{\rho}_2) = 2 - 2\,\mathrm{Tr}\left( \hat{\rho}_1 \hat{\rho}_2 \right) = 2 - 2 \left| \braket{\psi_1}{\psi_2} \right|^2.
\end{equation}
However $\hat{\rho}_1$ and $\hat{\rho}_2$ are in general mixed and, for given marginal distributions, $O_{\mathrm{pure}}(\hat{\rho}_1, \hat{\rho}_2)$ is not necessarily the minimum value that the operational distance can take. Hence it cannot be used to place any tighter bound on the mean purity.

\subsection{Apparatus and basic alignment}

\begin{figure}
\begin{center}
\includegraphics[width= 0.7 \textwidth]{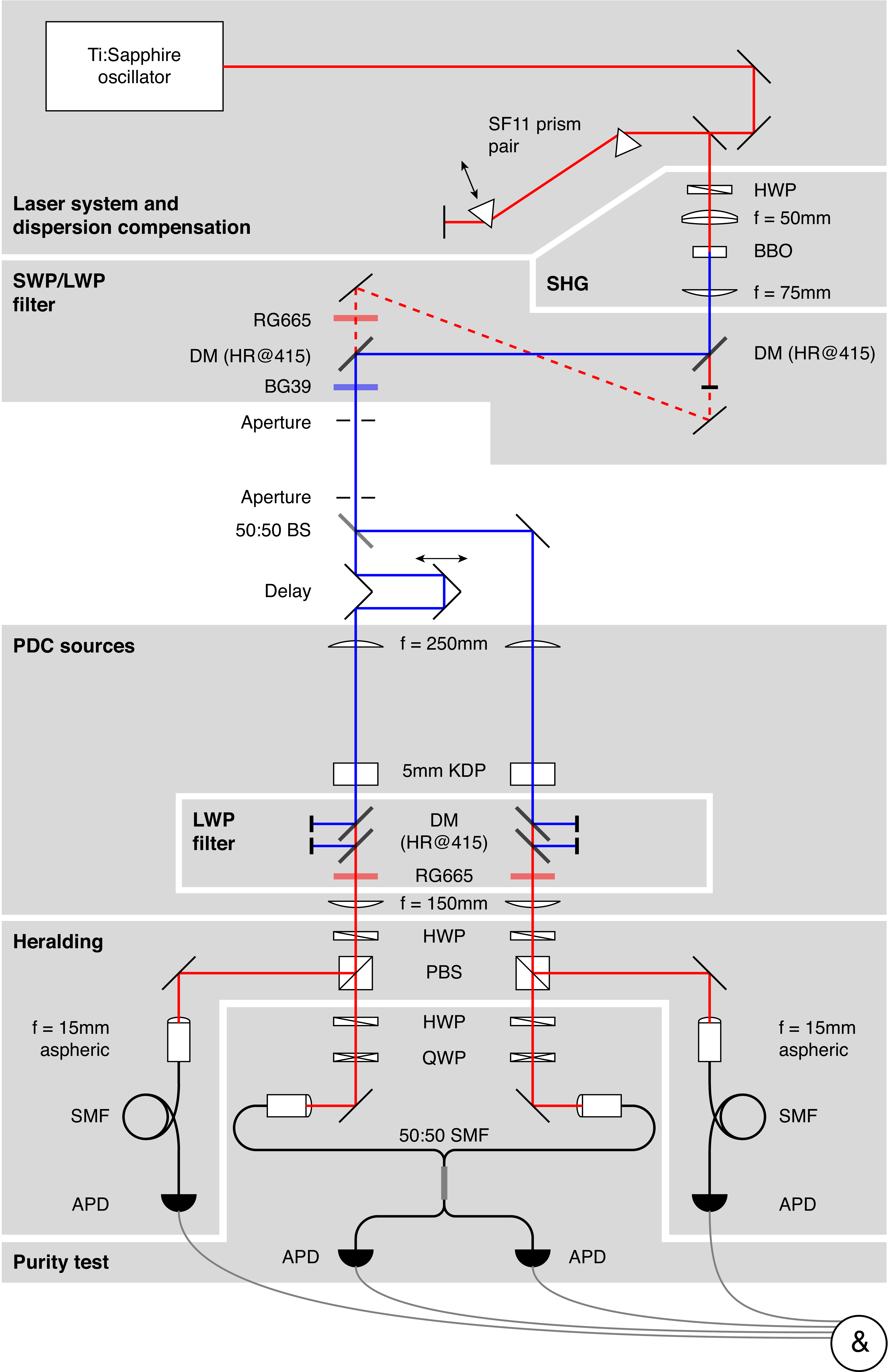}
\caption[Apparatus for two-crystal HOMI experiment]{Apparatus for two-crystal HOMI experiment. See text for details.}
\label{fig:homi_app}
\end{center}
\end{figure}

The experimental apparatus for the twin source HOMI experiment is displayed in figure \ref{fig:homi_app}. Pulses from a Ti:Sapphire oscillator (500\,mW, 76\,MHz, $\lambda_0$ = 830\,nm, 20\,nm FWHM) were first compressed in an external prism line before passing through a half waveplate (HWP) and then focused by an f = 50\,mm achromatic doublet into a 700\,$\mu$m BBO crystal angled for type-I phasematching at 830\,nm. This produced horizontally polarised second harmonic pulses centered at 415\,nm with an average power of around 150\,mW. The bandwidth at each point in the SHG beam was approximately 4\,nm and the chirp across the FWHM of the spatial distribution was 3.75\,nm in the horizontal direction and zero vertically. The frequency-doubled pulses were collimated with an f = 75\,mm fused silica (SiO$_2$) lens and the remaining fundamental beam filtered out with a combination of two dichroic mirrors (highly reflecting at 415\,nm and highly transmitting at 830\,nm) and a 2\,mm thick Schott BG39 short-wave-pass (SWP) coloured glass filter. The second harmonic power remaining after filtering, to use as the pump beam for the downconversion, was approximately 100\,mW. Upon moving a beam dump, an additional pair of dielectric mirrors (HR at 830\,nm) placed behind the dichroic mirrors, in conjunction with a Schott BG665 long-wave-pass (LWP) glass filter, allowed the remaining fundamental beam to be used as an alignment tool.

Two alignment apertures separated by 625\,mm preceded a 45$^{\circ}$ incidence 50:50 beamsplitter (BS) that divided the pump beam into two. Both beams were then directed to the two downconversion crystals, the reflected beam off one steering mirror and the transmitted beam via a time delay controlled by an automated translation stage to match the arrival times of the two pulses at the crystals. The pump beams, each with a power of 40\,mW, were both focused with f = 250\,mm SiO$_2$ lenses into two 5\,mm long KDP crystals cut for type-II phasematching placed one focal length away from the lenses. The KDP crystals were orientated with their optic axes in the horizontal plane and mounted in mirror mounts to allow fine control of the rotation angle about their vertical axes. Following the downconverters, the remaining pump light was filtered out firstly by a pair of SiO$_2$ LWP dichroic mirrors in each arm and secondly by 2\,mm thick anti-reflection (AR) coated RG665 LWP filters. Subsequently, the downconverted pairs were collimated with f = 150\,mm lenses.

The collimated pairs in each arm were then separated with polarising beamsplitters (PBS), each preceded by a HWP. Setting the HWPs not to rotate their input polarisation led to the e-ray photons being transmitted at the PBSs and the o-rays reflected, whereas rotating the polarisation of the pairs by 90$^{\circ}$ gave the opposite outcome at the PBSs. The reflected beams were then each coupled into single-mode fibre (SMF) using a steering mirror, an f = 15\,mm aspheric lens, and 3-axis flexure stages. The other end of these fibres went directly to two Perkin-Elmer SPCM silicon avalanche photodiodes (APDs) to act as the herald detectors. Each transmitted beam from the two PBSs passed first through a HWP and then a quarter waveplate (QWP) to allow pre-compensation of the polarisation rotation induced by the fibre pigtails preceding the 50:50 single-mode fibre coupler (also known as a fibre beamsplitter (FBS)); the fibre coupling arrangement here was the same as for the reflected arm. The two outputs of the 50:50 FBS went to another two Si APDs, the signal detectors. Coincidences between the signals from the APDs were monitored using electronics connected to a counting card interfaced with a PC. As required, the coincidence window was approximately 5\,ns --- substantially smaller than the 13\,ns between consecutive pulses from the laser system.

The downconversion crystals were initially aligned roughly using the fundamental Ti:Sapphire pulses. To manage properly the effects of spatial chirp, it was essential to mount both KDP crystals with their optic axes parallel and in the correct orientation with respect to the direction of the spatial chirp. Although manufacturers usually mark nonlinear crystals to indicate a plane in which the optic axis lies, in general they do not designate the direction of the axis. So it was with the KDP crystals used here: the plane of the optic axis was shown but it was still necessary to distinguish between the two possibilities for each of their directions. To this end, with the fundamental alignment beam incident on the crystals, a HWP was temporarily placed in the beam to rotate the polarisation to 45$^{\circ}$ and produce type-II SHG in the crystal. By placing a SWP filter and spectrometer after the crystal, the SHG spectrum was seen to shift as the crystal was rotated. In this way the orientation of the optic axes could be determined and the KDP crystals set in the correct direction to make the sign of the spatial chirp (as defined in section \ref{sec:modelling}) positive. Both crystals were then aligned to be at approximately normal incidence.

The count rates were then optimized with the second harmonic pulses from the BBO. For a second harmonic power of about 40\,mW per KDP crystal, the measured twofold coincidence rates were around 3000\,s$^{-1}$, suggesting a true coincidence rate (without the FBS in place) of 6000\,s$^{-1}$. This resulted in a fourfold coincidence rate of approximately 0.3\,s$^{-1}$ after the FBS, commensurate with the measured twofold rates and laser repetition frequency. Although the optical setup was fairly complex, once aligned it would remain reasonably stable over days and sometimes weeks at a time, requiring only small adjustments to keep it so.

\subsection{Mode-matching}

As no spectral filtering was used, it was vital to control directly the spectral mode overlap of the two heralded photons at the point of production through the parameters of the sources themselves, specifically their phasematching angles. Due to the high sensitivity of the e-ray wavelength to the phasematching angle, it was the e-ray spectrum that required the most care to fully mode match. With both crystals set to be at normal incidence, the two e-ray spectra were partially overlapped, though still offset by a significant fraction of their bandwidth. Two techniques were used to maximize the spectral overlap. The first was simply to make a direct measurement of the e-ray spectra using a spectrometer with a high-sensitivity CCD camera (in this case an Andor iXon electron-multiplying CCD). Adjusting the angle of the KDP crystals about the vertical axis (perpendicular to the principal plane) allowed the two e-ray spectra to be matched to within 0.1\,nm or so.

The second technique was to observe an interference effect in the twofold detection rates. With the HWPs before the PBSs set to rotate the polarisation of the e- and o-rays to $\pm$45$^{\circ}$, there was a 25\% chance that both photons from each pair would be transmitted towards the FBS. By reconfiguring the electronics to look for twofold coincidences between the outputs of the FBS (without heralding), the interference between the amplitude for creation of a pair in one crystal and creation of a pair in the other was observed as the time delay was scanned through zero. This interference between two possible (unheralded) paths involved the detection of only two photons, and therefore required only one pair to be generated per pump pulse (rather than two pairs in the case of the full HOMI interference). Hence the data could be monitored at each time delay setting for only a short time interval (0.5\,s) while still registering a significant number of counts. Furthermore, this interference effect makes no demands on the factorability of the joint spectral amplitude function from each crystal, only that they are identical to one another. Therefore it provided a perfect method of matching the spectral modes from  the two crystals.

\begin{figure}
\begin{center}
\includegraphics[width= 0.4 \textwidth]{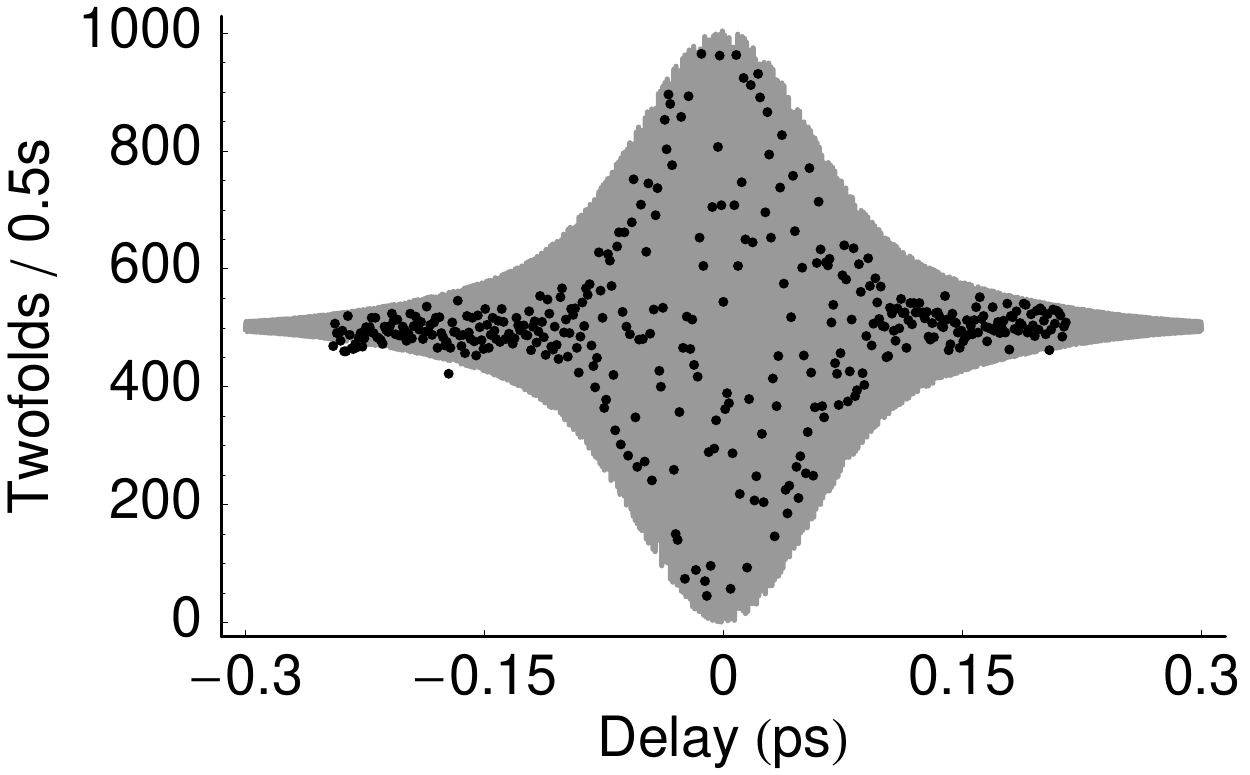}
\caption{Alignment interference between the amplitudes for creation of a pair in one crystal or the other. Solid area is a theoretical plot, points are data given the same parameters.}
\label{fig:alignment_interference}
\end{center}
\end{figure}

Real-time observation of this twofold interference effect allowed one crystal angle to be adjusted to maximize the interference visibility and once this optimization had taken place, the spectra from the two crystals were maximally overlapped. Note that the changes in angle required were fractions of a degree and hence the adjustments to the crystal angle made little difference to the optical time delay. The necessary range of angles could therefore be accessed without changing the zero point of the delay noticeably. The interference data displayed in figure \ref{fig:alignment_interference} were taken after this adjustment had been made.

\subsection{Outline of results}

First the interference of the two e-ray photons was studied by setting the HWPs to transmit these at the PBSs. The fourfold count rates were recorded for periods of 900\,s at a series of 13 equidistant delay stage positions over a range of about 200\,$\mu$m, corresponding to a total delay range of approximately 1.5\,ps. The twofold rates from each crystal were also recorded for periods of 1\,s at the same delay settings.

\begin{figure}
\begin{center}
\includegraphics[width= 0.9 \textwidth]{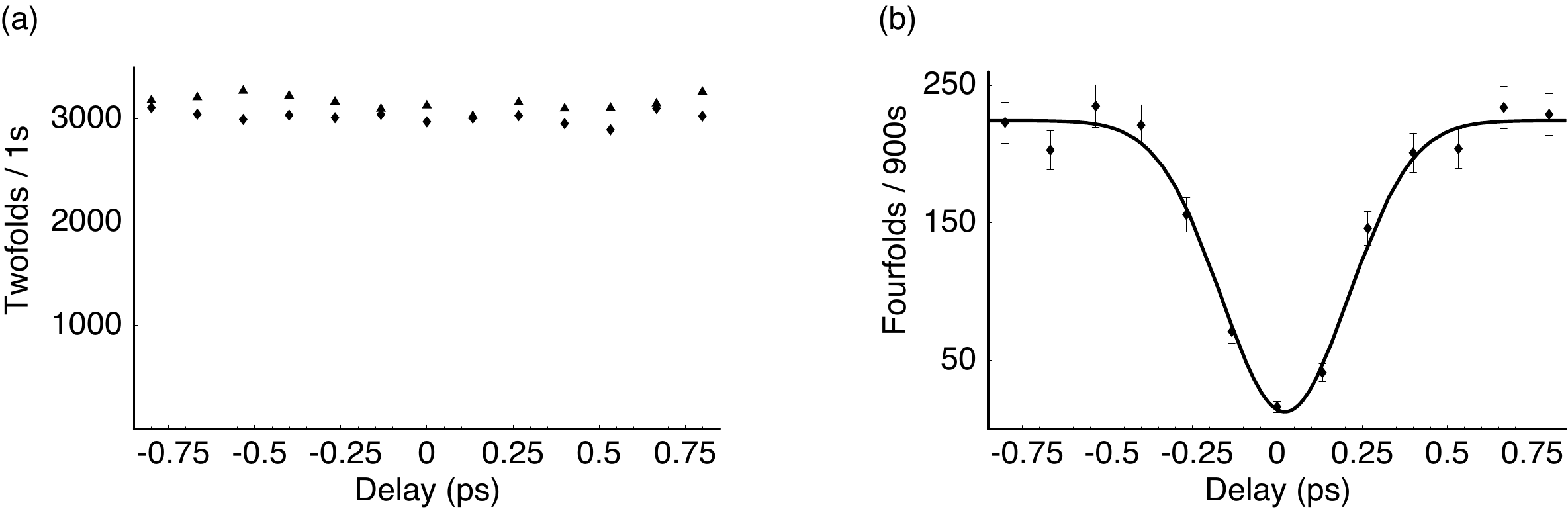}
\caption[Interference of the unfiltered heralded e-ray photons]{Interference of the unfiltered heralded e-ray photons. (a) Constant twofold coincidence rates. (b) Fourfold coincidence data for e-ray HOMI dip, displaying $V = 0.944 \pm 0.016$. The error bars, simply from Poissonian counting statistics, are equal to the square root of the number of counts at each point.}
\label{fig:eray_dip_data}
\end{center}
\end{figure}

The resulting raw data (without subtraction of background counts which were in any case minimal in the fourfold rates) are displayed in figure \ref{fig:eray_dip_data}. The fourfold coincidence data were fitted with a Gaussian dip function with four free parameters: the value for large delay, the visibility, the width, and the position of the zero delay point. The HOMI dip for the heralded e-ray photons from independent sources recorded without any spectral filters was thus found to have a visibility of $V = 0.944 \pm 0.016$ (standard error on the fit) and a FWHM of 440\,fs. The measured visibility agrees remarkably well with the value of 0.95 predicted by the numerical model. The temporal width of the dip corresponds to a coherence time of 310\,fs, commensurate with the measured spectral bandwidth of the e-ray photons --- a pulse at 830\,nm with 3.5\,nm of bandwidth would have a coherence time of 290\,fs.

\begin{figure}
\begin{center}
\includegraphics[width= 0.9 \textwidth]{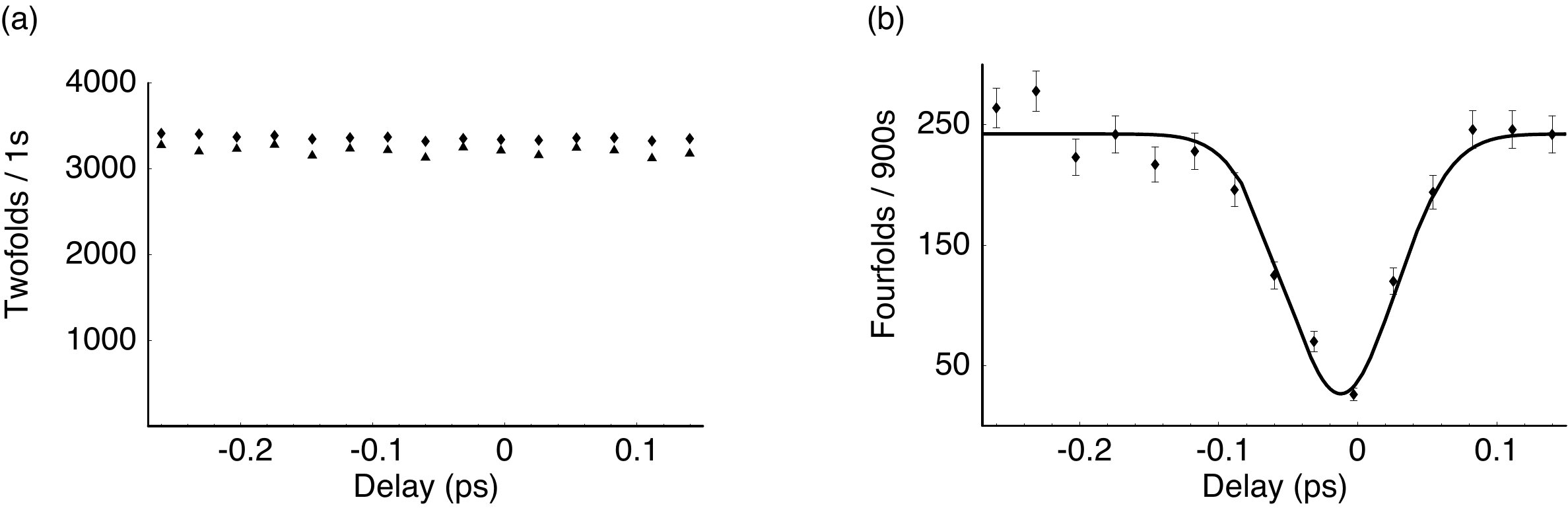}
\caption[Interference of the unfiltered heralded o-ray photons]{Interference of the unfiltered heralded o-ray photons. (a) Constant twofold coincidence rates. (b) Fourfold coincidence data for e-ray HOMI dip, displaying $V = 0.891 \pm 0.030$. Note much narrower temporal width than for the e-ray interference. The error bars represent Poissonian counting statistics.}
\label{fig:oray_dip_data}
\end{center}
\end{figure}

Secondly, the two o-ray photons were interfered by rotating the polarisation before the PBSs by 90$^{\circ}$. The fourfold coincidences were again recorded for 900\,s per point and are shown in Figure \ref{fig:oray_dip_data}. The visibility of the HOMI dip for the o-ray photons was $V = 0.891 \pm 0.030$ with a FWHM width of 92\,fs. This width gives a coherence time of approximately 65\,fs, also consistent with the broad measured spectral bandwidth (16.5\,nm bandwidth suggests a coherence time of 62\,fs). As there is little phase on the joint spectrum, shown by the high interference visibility, the time-bandwidth product of these photons should be almost Fourier transform limited, and their temporal duration will therefore only be slightly longer than their coherence time.

The detection efficiency, $\eta_D$, is the probability of finding a single photon in the signal arm given that the trigger has fired, and is defined as the ratio of coincidence detection events, $R_C$, at two detectors following a source to single detections at the trigger detector, $R_T$. The heralding efficiency of a source, $\eta_H$, is the probability that a single photon is delivered given that the trigger has fired and is related to the measured detection efficiency through the quantum efficiency of the detector in the signal arm, $\eta_q$:
\begin{equation}
\label{eq:detection_eff}
\eta_H = \eta_q \eta_D = \eta_q \frac{R_C}{R_T}.
\end{equation}
Knowledge of $\eta_q$ therefore allows an estimate to be made of the heralding efficiency of a source in the presence of imperfect detectors. During the data runs presented above the typical detection efficiency was around 6.5\%. However, in this case we must also take into account that the presence of the FBS in the signal arm reduced the number of coincidence events to half the value that would have been measured were the FBS replaced with a single length of fibre. The true detection efficiency at which the system was performing could then be inferred as twice the measured efficiency, or about 13\%.

However, when using the e-ray photons as heralds the asymmetry of the joint spectral distribution can be exploited to increase the heralding efficiency of the source by careful spectral filtering. By placing in the e-ray arm only a high-transmission filter whose bandwidth just exceeds that of the e-ray photons (around 4\,nm), many of the background counts (most of which come from the fundamental Ti:Sapphire beam that has a bandwidth around five times that of the e-ray photons) can be cut out without affecting the e-ray PDC photons. Therefore the singles rate in the herald arm can be reduced dramatically while not changing the form of the two-photon state or discarding any PDC photons.

To test this proposal, a spectral filter with a bandwidth of 3\,nm (slightly narrower than the ideal, but the only suitable filter available) and peak transmission well over 95\% at 830\,nm was placed in the e-ray arm of one of the downconversion sources configured to use the e-rays as heralds. The filter was angle tuned to set its central transmission wavelength to maximize the twofold coincidence rate from the source. In this way the detection efficiency was measured to be greater than 13\% with the FBS in place, giving a true detection efficiency of over 26\%. Furthermore, taking into account the quantum efficiency of the APD in the signal arm (about 60\% at this wavelength), the source heralding efficiency can be estimated at almost 44\%. Although higher heralding efficiencies have been measured from spectrally correlated PDC sources \cite{URen2004Efficient-Conditional-Preparation, Pittman2005Heralding-single-photons, Fedrizzi2007A-fiber-coupled-wavelength-tunable}, this is still very high for a pulsed bulk PDC source.

\subsection{Second source demonstration}

Since performing the initial successful implementation of group velocity matched downconversion described above, we subsequently built a second pair of sources almost identical to the first but pumped by an oscillator with higher power and narrower bandwidth (Spectra-Physics MaiTai HP, 2.8\,W, 80\,MHz, 10\,nm FWHM bandwidth). With approximately 350\,mW of frequency-doubled light incident on each KDP crystal the pair coincidence rates were $82\times10^3$\,s$^{-1}$ at a raw heralding efficiency of 18\%. This gave a mean fourfold coincidence rate of 35\,s$^{-1}$. This scaling is roughly consistent with the rate of collected photon pairs being proportional to the PDC pump power and therefore the rate of fourfolds scales with the square of the pump power. Measuring the HOMI dip with the e-ray photons from these two sources yielded an interference pattern with a raw visibility of $0.850 \pm 0.021$, significantly lower than for the first pair of sources. However, the data here contained a significant contribution from a constant background of fourfold detection events caused by simultaneous emission of two pairs from one crystal and one from the other. Therefore the drop in visibility was not a result of additional mixedness or distinguishability and hence the background contribution could be subtracted in order to reliably estimate the minimum mean photon purity. This was done by measuring the background from three-pair events as described in reference \cite{Fulconis2007Nonclassical-Interference-and-Entanglement} and subtracting this constant level from the fourfold counts, giving a corrected visibility of $0.950 \pm 0.023$. The construction of a second pair of sources capable of generating photons of similarly high purity as the original demonstration is important as it clearly shows that the group velocity matching technique in KDP is generalizable to other laser systems, even those without such broadband outputs. Furthermore, the exceptionally high fourfold coincidence rate obtainable from our new sources will allow experiments to move into the multi-photon, multi-source regime.

\section{Conclusion}

Through the observation of high-visibility Hong--Ou--Mandel interference between heralded single photons from two independent sources, factorable photon-pair generation by group velocity matching in KDP has been shown to be an efficient method of conditionally preparing high-purity single photons. However, in any experimentally feasible situation, one must be careful that during the optimization of the pumping and pair collection parameters to yield high count rates,  the conditions required for factorable state generation are maintained. As group velocity matching occurs in KDP for a pump wavelength of 415\,nm and an ultrafast pump is necessary to allow the phasematching function to dominate the form of the joint state, the downconversion crystals were pumped by a frequency-doubled Ti:sapphire laser. Spectral inhomogeneity across the pump pulses (spatial chirp) induced by tight focusing in the SHG crystal can either enhance or degrade the factorability of the photon pairs and must therefore be properly managed.

In this paper we have presented a numerical model that allows these effects to be simulated. By considering the contributions  to the joint state of noncollinear phasematching over the range of wavevectors contained in both the pump beam and collected pairs, the amplitude of the state coupled into single-mode fibres can be calculated. Hence, for a given set of pump beam parameters, focusing conditions can be found that maximize the purity of the heralded photons produced. Therefore, a source can be designed for a given ultrafast laser system that will provide both high count rates and high-visibility interference without the need for any spectral filtering. The predictions of the model have been compared with measurements of the joint spectral probability distribution and shown to be in good agreement. The heralded photons thus produced have been interfered with record visibility and shown ultrashort coherence times with almost Fourier transform limited spatiotemporal structures. We hope that this work will allow others to implement in the laboratory this very promising and robust source of photon pairs.

\section*{Acknowledgements}

This work was supported by the EPSRC (UK) through the QIP IRC (GR/S82716/01) and project EP/C013840/1 and by the European Commission under the Integrated Project Qubit Applications (QAP) funded by the IST directorate as Contract Number 015848. BJS acknowledges support from the Royal Society.

\begin{sidewaystable}[h]
\small{
\begin{tabular}{p{4.5cm} p{2.5cm} p{11cm}}		\hline
\textsf{Subsystem}				&	\textsf{Parameters}				&	\textsf{Comments}	\\ \hline

\textbf{Laser}				&	R\dotfill76\,MHz			& $\bullet$ Repetition frequency of standard modelocked oscillator	\\
\textit{Ti:Sapphire oscillator}	&	$\lambda_0$\dotfill830\,nm		& $\bullet$ Central wavelength for asymmetric group velocity matching in KDP	\\
						&	$\Delta \lambda$\dotfill20\,nm	&	$\bullet$ Broad spectral bandwidth of fundamental allows broadband output from frequency doubler	\\
						&	$\Delta \tau$\dotfill50\,fs			&	$\bullet$ Approximately transform limited pulses give high conversion efficiency to second harmonic and minimal temporal chirp on doubled pulses \\ \hline
					
\raggedright \textbf{SHG} \\ \textit{Type-I phasematched BBO}		&	L\dotfill700\,$\mu$m	&	$\bullet$ Short doubling crystal required to maintain broad spectral bandwidth in second harmonic	\tabularnewline
		&	$\lambda_p$\dotfill415\,nm	&	$\bullet$ Central pump wavelength to downconvert to group velocity matched wavelength: $v_{g,e}(415\,\mathrm{nm}) = v_{g,o}(830\,\mathrm{nm})$ in KDP	\\
								&	$\Delta \lambda$\dotfill4\,nm	&	\raggedright $\bullet$ Bandwidth at each point in beam at 415\,nm is broad to satisfy (\ref{eq:condition2}) \\ $\bullet$ Broadband pump function imposes minimal constraints on form of joint spectrum of photon pairs	\tabularnewline
								&	$\Delta \lambda_p$\dotfill3.75\,nm	&	$\bullet$ Spatial chirp across FWHM of SHG beam due to tight focusing required to give high conversion efficiency to second harmonic \\ \hline
		
\raggedright \textbf{PDC} \\ \textit{Type-II phasematched KDP}		&	L\dotfill5\,mm	&	\raggedright $\bullet$ Long crystal required to satisfy (\ref{eq:condition2}) \\ $\bullet$ Provides narrow phasematching function that dominates over broadband pump function giving ``vertical'' form to joint spectrum \\ $\bullet$ Reduces effects of any temporal phase of pump pulses on temporal structure of photon pairs \tabularnewline
		&	$\Delta \theta_{\text{in}}$\dotfill0.16$^\circ$	&	\raggedright $\bullet$ FWHM range of input angles controls spread of phasematching conditions experienced by rays comprising pump beam \\ $\bullet$ Must be small to maintain uncorrelated joint two-photon state \\ $\bullet$ KDP must be orientated such that frequency change due to spatial chirp of pump is positively correlated with phasematching angle \\ $\bullet$ Change in pump function across beam counteracts changing phasematching conditions \\ $\bullet$ Spatial chirp in principal plane of crystal \tabularnewline
			&	$\Delta \theta_{\text{out}}$\dotfill0.30$^\circ$	&	$\bullet$ FWHM range of output angles also small to give uncorrelated state	\\ \hline
\end{tabular}}
\caption{Critical source parameters.}
\label{tab:details}
\end{sidewaystable}

\end{document}